\documentclass[12pt]{article}

\usepackage{graphicx}
\usepackage[center,footnotesize,hang]{subfigure}
\usepackage{amsmath,amssymb}
\usepackage{bm}
\usepackage{graphicx, color}
\usepackage{wrapfig}
\begin{document}
\title{
\begin{flushright}
\ \\*[-80pt] 
\begin{minipage}{0.2\linewidth}
\normalsize
\end{minipage}
\end{flushright}
{\Large \bf  Slepton Mass Matrices,  
$\mu\rightarrow e\gamma$ Decay and EDM
in SUSY $S_4$ Flavor Model
\\*[20pt]}}

\author{
\centerline{
Hajime~Ishimori$^{1,}$\footnote{E-mail address: ishimori@muse.sc.niigata-u.ac.jp} \quad   and \  
~Morimitsu~Tanimoto$^{2,}$\footnote{E-mail address: tanimoto@muse.sc.niigata-u.ac.jp} }
\\*[20pt]
\centerline{
\begin{minipage}{\linewidth}
\begin{center}
$^1${\it \normalsize
Graduate~School~of~Science~and~Technology,~Niigata~University, \\ 
Niigata~950-2181,~Japan } \\
$^2${\it \normalsize
Department of Physics, Niigata University,~Niigata 950-2181, Japan } 
\end{center}
\end{minipage}}
\\*[50pt]}

\date{
\centerline{\small \bf Abstract}
\begin{minipage}{0.9\linewidth}
\medskip 
\medskip 
\small
We discuss  slepton mass matrices 
in the   $S_4$ flavor model with SUSY $SU(5)$ GUT.
By considering  the gravity mediation within the 
framework of supergravity theory, 
we estimate  the SUSY breaking terms in the slepton mass matrices,
which contribute to the  $\mu \rightarrow e + \gamma$ decay.
 We  obtain 
a lower bound for the ratio of $\mu\rightarrow e\gamma$ as $10^{-13}$
 if $m_{\text{SUSY}}$ and  $m_{1/2}$ are below $500$GeV.
 The off diagonal terms of slepton mass matrices
also contribute to   EDM of leptons.
 The predicted electron EDM is around $10^{-29}-10^{-28}e$cm.
Our predictions are expected to be tested  in the near future experiments.
\end{minipage}
}

\begin{titlepage}
\maketitle
\thispagestyle{empty}
\end{titlepage}

\section{Introduction}

Recent experiments of the neutrino oscillation 
go into a  new  phase  of precise  determination of
 mixing angles and mass squared  differences
~\cite{Schwetz:2008er,Fogli:2008jx,Fogli:2009zza,GonzalezGarcia:2010er}, 
which indicate the tri-bimaximal mixing  for three flavors 
 in the lepton sector~\cite{Harrison:2002er,Harrison:2002kp,Harrison:2003aw,Harrison:2004uh}. 
These large  mixing angles are completely  
 different from the quark mixing ones.
Therefore, there appear many researches
to find a natural model that leads to the mass spectrum and mixing
 of quarks and leptons.
The flavor symmetry is expected to explain them.
In particular, the non-Abelian discrete symmetry of flavors
\cite{Ishimori:2010au}
has been studied  intensively in the quark and lepton sectors.
 Actually, the tri-bimaximal mixing of leptons has been at first understood 
based on the non-Abelian finite group $A_4$~\cite{Ma:2001dn,Ma:2002ge,Ma:2004zv,Altarelli:2005yp,Altarelli:2005yx}.
Until now,  much progress has been made in the  theoretical
and phenomenological  analysis of $A_4$ flavor model
~\cite{Babu:2002in}-\cite{Hirsch:2010ru}.
The other  attractive candidate of the flavor symmetry 
 is the $S_4$ symmetry, which
 was used for the neutrino masses and the neutrino flavor mixing
~\cite{Yamanaka:1981pa,Brown:1984dk,Brown:1984mq,Ma:2005pd}.
The exact tri-bimaximal neutrino mixing is realized
 in  $S_4$ flavor models~\cite{Lam:2008sh,Bazzocchi:2008ej,Ishimori:2008fi,Grimus:2009pg,Bazzocchi:2009pv,Bazzocchi:2009da,Meloni:2009cz}.
Many studies in the  $S_4$ flavor model
have been presented  for  quarks as well as   leptons
~\cite{Zhang:2006fv}-\cite{Ishimori:2010fs}.
Some works  attempt to unify the quark and lepton sectors
 toward a grand unified theory in the framework 
of the $S_4$ flavor symmetry~\cite{Hagedorn:2006ug,Cai:2006mf,Caravaglios:2005gw},
  however, quark mixing angles were not predicted clearly.

Recently, $S_4$ flavor models to unify
 quarks and leptons have been proposed
in the framework of the $SU(5)$ SUSY GUT
~\cite{Ishimori:2008fi} or  $SO(10)$  SUSY GUT 
\cite{Dutta:2009bj,Patel:2010hr}.
There also appeared the $S_4$ flavor model in $SU(5)$ SUSY GUT 
 ~\cite{Hagedorn:2010th,Ding:2010pc,Ishimori:2010xk} 
and the Pati-Salam SUSY GUT
\cite{Toorop:2010yh,Toorop:2010zg},
taking account of the next-to-leading order of  mass operators.
These unified models seem to explain both mixing of quarks and leptons.

Since many flavor models have been proposed,
 it is important to study how to test them.
The flavor symmetry  in the framework of  SUSY controls 
 the slepton and squark mass matrices as well as 
 the quark and lepton ones.
For example, the predicted slepton mass matrices reflect
 structures of  the charged  lepton mass matrix.
Therefore, the slepton mass matrices provide  us an important
 test for the flavor symmetry.

Our $S_4$ flavor model \cite{Ishimori:2010xk} is an attractive one because
 it gives the proper quark flavor mixing angles
  as well as the tri-bimaximal mixing of neutrino flavors.
 Especially, the Cabibbo angle is predicted to be  $15^\circ$
due to $S_4$ Clebsch-Gordan coefficients  in the leading order.
Including the next-to-leading corrections of the $S_4$ symmetry, the predicted   Cabibbo angle is completely consistent with the observed one.

 In our $S_4$ flavor model, three generations of $\overline 5$-plets 
in $SU(5)$ are assigned to ${\bf 3}$ of $S_4$ 
while the  first and second generations of 
$10$-plets  in  $SU(5)$  are assigned to ${\bf 2}$ of $S_4$,
and the third generation of $10$-plet is  assigned to ${\bf 1}$ of $S_4$.
These  assignments of $S_4$ for $\overline 5$ and $10$ 
lead to the  completely different structure 
of  quark and lepton mass matrices.
Right-handed neutrinos, which are $SU(5)$ gauge singlets, 
are also assigned to  ${\bf 2}$ for the first and second generations,
and ${\bf 1}'$ for  the third generation.
These  assignments realize the tri-bimaximal mixing
of neutrino flavors.

We discuss  slepton mass matrices 
in our $S_4$ flavor model
by considering  the gravity mediation within the 
framework of supergravity theory. 
We estimate the SUSY breaking in the slepton mass matrices
by taking account of  the next-to-leading 
$S_4$ invariant mass operators.
Then,  we can predict  the lepton flavor violation (LFV), e.g.,
 the  $\mu \rightarrow e + \gamma$ decay.
A similar study  of the LFV has been presented 
in the $A_4$ flavor model \cite{Feruglio}.
Slepton mass matrices also  give  the electric dipole moment (EDM) 
of the lepton \cite{Hisano:2007cz,Hisano:2008hn},
which has not been  discussed  in flavor models 
with the non-Abelian discrete symmetry.
 We predict the EDM of the electron 
 versus  the $\mu \rightarrow e + \gamma$ decay ratio,
which are important to study the SUSY sector comprehensively
\cite{Gabbiani:1996hi,Altmannshofer,Hisano:2009ae}.


In section 2,
we summarize briefly the $S_4\times Z_4\times U(1)_{FN}$ flavor model of
 quarks and leptons in  $SU(5)$ SUSY GUT including
    the higher dimensional mass operators.  
In section 3, the slepton mass matrices  are discussed precisely.
The numerical predictions of LFV processes and lepton EDM's
 are presented in section 4.
Section 5 is devoted to the summary.
The multiplication rule of $S_4$ is presented
in Appendix.

\section{Overview of $S_4$ flavor model with $SU(5)$ SUSY GUT}

 In this section, we summarize our $S_4$ flavor model \cite{Ishimori:2010xk}
 to unify quarks and leptons in the framework of the $SU(5)$ SUSY GUT. 
The $S_4$ group has 24 distinct elements and irreducible representations 
${\bf 1},~{\bf 1}',~{\bf 2},~{\bf 3}$, and ${\bf 3}'$, 
which are assigned for each  $SU(5)$ representation.\\

\begin{table}[t]
\begin{tabular}{|c|ccccc||cccc|}
\hline
&$(T_1,T_2)$ & $T_3$ & $( F_1, F_2, F_3)$ & $(N_e^c,N_\mu ^c)$ & $N_\tau ^c$ & $H_5$ &$H_{\bar 5} $ & $H_{45}$ & $\Theta $ \\ \hline
$SU(5)$ & $10$ & $10$ & $\bar 5$ & $1$ & $1$ & $5$ & $\bar 5$ & $45$ & $1$ \\
$S_4$ & $\bf 2$ & $\bf 1$ & $\bf 3$ & $\bf 2$ & ${\bf 1}'$ & $\bf 1$ & $\bf 1$ & $\bf 1$ & $\bf 1$ \\
$Z_4$ & $-i$ & $-1$ & $i$ & $1$ & $1$ & $1$ & $1$ & $-1$ & $1$ \\
$U(1)_{FN}$ & $1 $ & 0 & 0 & $1$ & 0 & 0 & 0 & 0 & $-1$ \\
\hline
\end{tabular}
\end{table}
\vspace{-0.5cm}
\begin{table}[t]
\begin{tabular}{|c|ccccccc|}
\hline
& $(\chi _1,\chi _2)$ & $(\chi _3,\chi _4)$ & $(\chi _5,\chi _6,\chi _7)$ 
& $(\chi _8,\chi _9,\chi _{10})$ & $(\chi _{11},\chi _{12},\chi _{13})$ & $\chi _{14}$ & $(\chi _{15},\chi _{16},\chi _{17})$ \\ \hline
$SU(5)$ & $1$ & $1$ & $1$ & $1$ & $1$ & $1$ & $1$ \\
$S_4$ & $\bf 2$ & $\bf 2$ & ${\bf 3}'$ & $\bf 3$ & $\bf 3$ & $\bf 1$& $\bf 3$ \\
$Z_4$ & $-i$ & $1$ & $-i$ & $-1$ & $i$ & $i$& $-1$ \\
$U(1)_{FN}$ & $-1 $ & $-2$ & 0 & 0 & 0 & $-1 $& $-z$ \\
\hline
\end{tabular}
\caption{Assignments of $SU(5)$, $S_4$, $Z_4$, and $U(1)_{FN}$ representations.}
\label{tables4}
\end{table}

In $SU(5)$, matter fields are unified into $10$ 
and $\bar 5$ dimensional representations. 
Three generations of $\bar 5$, which are denoted by $F_i~(i=1,2,3)$,
 are assigned to $\bf 3$ of $S_4$. 
On the other hand, the third generation of the $10$-dimensional 
representation, $T_3$, is assigned to $\bf 1$ of $S_4$, and  
 the first and second generations of $10$, $(T_1, T_2)$, 
are assigned to $\bf 2$ of $S_4$, respectively. 
Right-handed neutrinos, which are $SU(5)$ gauge singlets, 
are also assigned to ${\bf 2}$ for the first and second generations,
$(N_e^c, N_\mu^c)$,
and ${\bf 1}'$ for  the third one, $N_\tau^c$. 
The $5$-dimensional, 
$\bar 5$-dimensional,  and $45$-dimensional Higgs of $SU(5)$, $H_5$, 
$H_{\bar 5} $, and $H_{45} $ are  assigned to $\bf 1$ of $S_4$. 
In order to obtain desired mass matrices, we introduce 
 $SU(5)$ gauge singlets $\chi_i$, so called flavons,  
which couple to quarks and leptons.


The $Z_4$ symmetry is added to obtain relevant couplings.
The Froggatt-Nielsen mechanism~\cite{Froggatt:1978nt}
  is  introduced  to get the natural hierarchy among quark and lepton masses,
as an additional 
$U(1)_{FN}$ flavor symmetry, where 
$\Theta$ denotes the Froggatt-Nielsen flavon.
The particle assignments of $SU(5)$, $S_4$, $Z_4$, and $U(1)_{FN}$
 are presented  in Table 1. 

The couplings of flavons are restricted as follows.
In the leading order, $(\chi _3,\chi _4)$ are  
coupled with the right-handed Majorana neutrino sector, 
$(\chi _5,\chi _6,\chi _7)$ are coupled with the Dirac neutrino sector, 
$(\chi _8,\chi _9,\chi _{10})$ and $(\chi _{11},\chi _{12},\chi _{13})$ 
are coupled with the charged lepton and down-type quark sectors.
  In the next-to-leading order, 
$(\chi_1,\chi_2)$ are coupled with the  up-type  quark sector, 
and  $\chi_{14}$ contributes 
 to the charged lepton and down-type quark sectors,
 and then the mass ratio of the electron and down quark is reproduced
properly.
The $S_4$ triplet  $(\chi _{15},\chi _{16},\chi _{17})$ 
does not couple with quarks and leptons 
directly due to $U(1)_{FN}$ as far as $z\gg 1$, but couples with other flavons to give alignments of vacuum expectation values (VEV's) as discussed later.

Our model predicts the quark  mixing  as well as the tri-bimaximal
mixing of leptons. Especially, the Cabibbo angle is
predicted to be     $15^{\circ}$ in the leading order.
The model is consistent with   the observed CKM mixing angles
and $CP$ violation
 as well as the non-vanishing $U_{e3}$ of the neutrino flavor mixing.


 Let us write down  the superpotential 
respecting  $S_4$, $Z_4$ and $U(1)_{FN}$
symmetries
 in terms of the $S_4$ cutoff scale $\Lambda$,  and
the $U(1)_{FN}$ cutoff scale  $\overline \Lambda$. 
In our calculation, 
both cutoff scales are taken as the GUT scale which is around $10^{16}$GeV. 
The $SU(5)$ invariant superpotential 
of the Yukawa  sector up to the linear terms of $\chi_i$ ($i=1,\cdots ,13$) is given as
\begin{align}
w &= y_1^u(T_1,T_2)\otimes T_3\otimes (\chi _1,\chi _2)\otimes H_5/\Lambda + y_2^uT_3\otimes T_3\otimes H_5 \nonumber \\
&\ + y_1^N(N_e^c,N_\mu ^c)\otimes (N_e^c,N_\mu ^c)\otimes \Theta ^{2}/\bar \Lambda  \nonumber \\
&\ + y_2^N(N_e^c,N_\mu ^c)\otimes (N_e^c,N_\mu ^c)\otimes (\chi _3,\chi _4) + MN_\tau ^c\otimes N_\tau ^c \nonumber \\
&\ + y_1^D(N_e^c,N_\mu ^c)\otimes (F_1,F_2,F_3)\otimes (\chi _5,\chi _6,\chi _7)\otimes H_5\otimes \Theta /(\Lambda \bar \Lambda ) \nonumber \\
&\ + y_2^DN_\tau ^c\otimes (F_1,F_2,F_3)\otimes (\chi _5,\chi _6,\chi _7)\otimes H_5/\Lambda \nonumber \\
&\ + y_1(F_1,F_2,F_3)\otimes (T_1,T_2)\otimes (\chi _8,\chi _9,\chi _{10})\otimes H_{45}\otimes \Theta /(\Lambda \bar \Lambda) \nonumber \\
&\ + y_2(F_1,F_2,F_3)\otimes T_3\otimes (\chi _{11},\chi _{12},\chi _{13})\otimes H_{\bar 5}/\Lambda ,
\end{align}
where $y_1^u$, $y_2^u$, $y_1^N$, $y_2^N$, $y_1^D$, $y_2^D$, 
$y_1$, and $y_2$ are Yukawa couplings of order one, 
and $M$ is the right-handed Majorana mass, which is taken to be $10^{12}$GeV 
in our calculation. 
 We can discuss the feature of the quark and lepton mass matrices 
 and flavor mixing based on this superpotential 
by using the $S_4$ multiplication rule in Appendix.


We require vacuum alignments for the VEV's of flavons in order to
get desired quarks and leptons mass matrices. 
The alignment depends on the structure of the scalar potential 
which is constructed by adding driving fields
$\chi_1^0$, $\chi_2^0$,  $\chi_3^0$ and $(\chi_4^0, \chi_5^0)$
 with having  $U(1)_R$ charge two as shown in Table 2. 
Matter fields ($T_i$, $F_i$, and $N_i$) are assigned to $U(1)_R$ charge one 
and Higgs, flavons are assigned to zero. 
A continuous $U(1)_R$ symmetry contains the usual R-parity as a
subgroup. 
 
The superpotential of the scalar sector including driving fields is given by 
\begin{align}
w' &= \kappa _1\left (\chi _1,\chi _2\right )\otimes \left (\chi _1,\chi _2\right )\otimes \left (\chi _3,\chi _4\right )\otimes \chi _1^0/\Lambda \nonumber \\
&\ +\eta _1\left (\chi _8,\chi _9,\chi _{10}\right )\otimes \left (\chi _{11},\chi _{12},\chi _{13}\right )\otimes \chi _2^0 \nonumber \\
&\ +\eta _2\left (\chi _1,\chi _2\right )\otimes \left (\chi _1,\chi _2\right )\otimes \chi _3^0+\eta _3\chi _{14}\otimes \chi _{14}\otimes \chi _3^0 \nonumber \\
&\ +\eta _4\left (\chi _5,\chi _6,\chi _7\right )\otimes \left (\chi _{15},\chi _{16},\chi _{17}\right )\otimes \left (\chi _4^0,\chi _5^0\right ),
\label{scalar-leading}
\end{align}
where $\kappa_1$ and $\eta_i$ ($i=1$--4) are coupling constants of order one. 
It gives the scalar potential
\begin{align}
V &= \left |\frac{\kappa _1}{\Lambda}
\left [2\chi _1\chi _2\chi _3+\left (\chi _1^2-\chi _2^2\right )\chi _4\right ] \right |^2 
 + \left |\eta _1\left (\chi _8\chi _{11}+\chi _9\chi _{12}+\chi _{10}\chi _{13}\right )\right |^2 \nonumber \\
&\ + \left |\eta _2(\chi _1^2+\chi _2^2)+\eta _3\chi _{14}^2\right |^2 \nonumber 
+ \left |\frac{1}{\sqrt 2}\eta _4\left (\chi _6\chi _{16}-\chi _7\chi _{17}\right )\right |^2 \nonumber \\
&\ + \left |\frac{1}{\sqrt 6}\eta _4\left (-2\chi _5\chi _{15}+\chi _6\chi _{16}+\chi _7\chi _{17}\right )\right |^2 \ .
\end{align}
Therefore, conditions to realize  the potential minimum ($V=0$) 
are given  as 
\begin{eqnarray}
&& (\chi _1, \chi _2)= (1,1), \quad 
 (\chi _3, \chi _4)= (0,1), \quad
 (\chi _5, \chi _6, \chi _7)=(1,1,1),
 \quad (\chi _8, \chi _9, \chi _{10})=(0,1,0), 
\nonumber\\
&&(\chi _{11}, \chi_{12}, \chi _{13})=(0,0,1), 
\quad \chi_{14}^2=-\frac{2\eta_2}{\eta_{3}}\chi_1^2,
\quad (\chi_{15},\chi_{16},\chi_{17})=(1,1,1), 
\label{alignment}
\label{vev}
\end{eqnarray}
where  these magnitudes are given in  arbitrary units.
Hereafter, we suppose these gauge-singlet scalars 
develop VEV's by denoting $\langle \chi_i\rangle=a_i\Lambda$, 
where $a_i$'s are given to be same order as shown in section 4.  

Denoting Higgs doublets as $h_u$
and $h_d$, we take VEV's of following scalars by
\begin{eqnarray}
\langle h_u\rangle 
= v_u, 
\quad
\langle h_d\rangle 
= v_d, 
\quad
\langle h_{45}\rangle 
= v_{45}, 
\quad 
\langle\Theta\rangle =\theta ,
\end{eqnarray}
which are supposed to be real. 
We define $\lambda = \theta/\Lambda$ 
to describe the Froggatt-Nielsen mechanism.

\begin{table}[t]
\begin{center}
\begin{tabular}{|c|cccc|}
\hline
& $\chi _1^0$ & $\chi _2^0$ & $\chi _3^0$ & $(\chi _4^0,\chi _5^0)$ \\ \hline
$SU(5)$ & $1$ & $1$ & $1$ & $1$ \\
$S_4$  & $\bf 1$ & $\bf 1$ & $\bf 1$ & $\bf 2$ \\
$Z_4$  & $-1$ & $i$ & $-1$ & $-i$ \\
$U(1)_{FN}$  & $4$ & $0$ & $2 $ & $z$ \\
$U(1)_R$  & $2$ & $2$ & $2$ & $2$ \\
\hline
\end{tabular}
\caption{Assignments of $SU(5)$, $S_4$, $Z_4$, and $U(1)_{FN}$ representations.}
\label{tables4}
\end{center}
\end{table}


First we consider mass matrices of the lepton sector. 
Taking vacuum alignments in  Eq. (\ref{vev}), 
the mass matrix of charged lepton becomes 
\begin{equation}
M_l = \begin{pmatrix}
                                   0 & -3y_1\lambda  a _9v_{45}/\sqrt 2 & 0 \\
                                   0 & -3y_1\lambda  a _9v_{45}/\sqrt 6 & 0 \\
                                   0 & 0 & y_2a _{13}v_d
                                \end{pmatrix},
\end{equation}
then, masses are given as
\begin{align}
m_e^2 = 0 \ ,
\quad
m_\mu ^2 =6|\bar y_1\lambda  a _9|^2v_d^2\ ,
\quad 
m_\tau ^2=|y_2|^2a _{13}^2v_d^2\ .
\label{chargemass}
\end{align}
In the same way, 
the right-handed Majorana mass matrix of neutrinos is given by 
\begin{equation}
M_N = \begin{pmatrix}
               y_1^N\lambda ^2\bar \Lambda +y_2^Na _4\Lambda & 0 & 0 \\
               0 & y_1^N\lambda ^2\bar \Lambda -y_2^Na _4\Lambda & 0 \\
               0 & 0 & M
         \end{pmatrix},
\end{equation}
and the Dirac mass matrix of neutrinos is
\begin{equation}
M_D = y_1^D\lambda v_u\begin{pmatrix}
        2a _5/\sqrt 6 & -a _5/\sqrt 6 & -a _5/\sqrt 6 \\
           0 & a _5/\sqrt 2 & -a _5/\sqrt 2 \\
                                        0 & 0 & 0 
                                     \end{pmatrix}+y_2^Dv_u\begin{pmatrix}
                                     0 & 0 & 0 \\
                                     0 & 0 & 0 \\
                         a _5 & a _5 & a _5
                                                \end{pmatrix}.
\end{equation}
By using the seesaw mechanism $M_\nu = M_D^TM_N^{-1}M_D$, 
the left-handed Majorana neutrino mass matrix is  written as
\begin{equation}
M_\nu = \begin{pmatrix}
                 a+\frac{2}{3}b & a-\frac{1}{3}b & a-\frac{1}{3}b \\
 a-\frac{1}{3}b& a+\frac{1}{6}b+\frac{1}{2}c & a+\frac{1}{6}b-\frac{1}{2}c \\
 a-\frac{1}{3}b & a+\frac{1}{6}b-\frac{1}{2}c & a+\frac{1}{6}b+\frac{1}{2}c
            \end{pmatrix},
\label{neutrino}
\end{equation}
where
\begin{equation}
a = \frac{(y_2^Da _5v_u)^2}{M},\qquad 
b = \frac{(y_1^Da _5v_u\lambda)^2}{y_1^N\lambda ^2\bar \Lambda +y_2^Na _4\Lambda},\qquad 
c = \frac{(y_1^Da _5v_u\lambda)^2}{y_1^N\lambda ^2\bar \Lambda -y_2^Na _4\Lambda}.
\label{neutrinomassparameter}
\end{equation}
It gives the tri-bimaximal mixing matrix 
$U_\text{tri-bi}$ and mass eigenvalues  as follows:
\begin{eqnarray}
&&U_\text{tri-bi} = \begin{pmatrix}
               \frac{2}{\sqrt{6}} &  \frac{1}{\sqrt{3}} & 0 \\
     -\frac{1}{\sqrt{6}} & \frac{1}{\sqrt{3}} &  -\frac{1}{\sqrt{2}} \\
      -\frac{1}{\sqrt{6}} &  \frac{1}{\sqrt{3}} &   \frac{1}{\sqrt{2}}
         \end{pmatrix},
\nonumber\\
\nonumber\\
&& m_{\nu_1} = b\ ,\qquad m_{\nu_2} = 3a\ ,\qquad m_{\nu_3} = c\ .
\label{mass123}
\end{eqnarray}
 The next-to-leading terms of the superpotential are important
to predict the deviation from the tri-bimaximal mixing of leptons,
especially, $U_{e3}$.  
 The relevant superpotential in the charged lepton sector
 is given at the next-to-leading order  as 
\begin{align}
\Delta w_l&=y_{\Delta _a}(T_1,T_2)\otimes (F_1,F_2,F_3)\otimes (\chi _1,\chi _2)\otimes (\chi _{11},\chi _{12},\chi _{13})\otimes H_{\bar 5}/\Lambda ^2 \nonumber \\
&\ +y_{\Delta _b}(T_1,T_2)\otimes (F_1,F_2,F_3)\otimes (\chi _5,\chi _6,\chi _7)\otimes \chi _{14}\otimes H_{\bar 5}/\Lambda ^2 \nonumber \\
&\ +y_{\Delta _c}(T_1,T_2)\otimes (F_1,F_2,F_3)\otimes (\chi _1,\chi _2)\otimes (\chi _5,\chi _6,\chi _7)\otimes H_{45}/\Lambda ^2 \nonumber \\
&\ +y_{\Delta _d}(T_1,T_2)\otimes (F_1,F_2,F_3)\otimes (\chi _{11},\chi _{12},\chi _{13})\otimes \chi _{14}\otimes H_{45}/\Lambda ^2 \nonumber \\
&\ +y_{\Delta _e}T_3\otimes (F_1,F_2,F_3)\otimes (\chi _5,\chi _6,\chi _7)\otimes (\chi _8,\chi _9,\chi _{10})\otimes H_{\bar 5}\otimes /\Lambda ^2 \nonumber \\
&\ +y_{\Delta _f}T_3\otimes (F_1,F_2,F_3)\otimes (\chi _8,\chi _9,\chi _{10})\otimes (\chi _{11},\chi _{12},\chi _{13})\otimes H_{45}\otimes /\Lambda ^2\ .
\label{nextsusy}
\end{align} 
By using this superpotential,
 we obtain the charged lepton mass matrix as
\begin{equation}
M_l\simeq
\begin{pmatrix}
\epsilon _{11} & \frac{\sqrt 3m_\mu}{ 2}+\epsilon_{12} & \epsilon _{13} \\
\epsilon _{21} & \frac{m_\mu}{2}+\epsilon_{22} & \epsilon _{23} \\
\epsilon _{31} & 0 & m_\tau+\epsilon_{33}
\end{pmatrix},
\label{nextleading}
\end{equation}
where $m_\mu$ and $m_\tau$ are given in  Eq.~(\ref{chargemass}),
and $\epsilon_{ij}$'s  are calculated by using  Eq.~(\ref{nextsusy})
to find
\begin{align}
\epsilon_{11}&=y_{\Delta _b}a _5a _{14}v_d
   -3\bar y_{\Delta _{c_2}}a _1a _5v_d , \nonumber \\
\epsilon_{12}&=
   -\frac{1}{2}y_{\Delta _b}a _5a _{14} v_d   
   +3\left [  \frac{\sqrt 3}{4}(\sqrt 3-1)\bar y_{\Delta _{c_1}}-\frac{1}{4}(\sqrt 3+1)\bar y_{\Delta _{c_2}}  
   \right ]a _1a _5v_d , \nonumber \\
\epsilon_{13}&=\left [\left \{ \frac{\sqrt 3}{4}(\sqrt 3-1)y_{\Delta _{a_1}}+\frac{1}{4}(\sqrt 3+1)y_{\Delta _{a_2}}\right \} a _1a _{13} 
   -\frac{1}{2}y_{\Delta _b}a _5a _{14}\right ]v_d  \nonumber \\
   &\ -3\left [\left \{ -\frac{\sqrt 3}{4}(\sqrt 3+1)\bar y_{\Delta _{c_1}}-\frac{1}{4}(\sqrt 3-1)\bar y_{\Delta _{c_2}}\right \}a _1a _5
   +\frac{\sqrt 3}{2}\bar y_{\Delta _d}a _{13}a _{14}\right ]v_d , \nonumber \\
\epsilon_{21}&=-3 \bar y_{\Delta _{c_1}}a _1a _5v_d , \nonumber \\
\epsilon_{22}&=\frac{\sqrt 3}{2}y_{\Delta _b}a _5a _{14} v_d
   +3\left [ \frac{1}{4}(\sqrt 3-1)\bar y_{\Delta _{c_1}}
   +\frac{\sqrt 3}{4}(\sqrt 3+1)\bar y_{\Delta _{c_2}}  
   \right ]a _1a _5 v_d , \nonumber \\
\epsilon_{23}&=\left [\left \{ -\frac{1}{4}(\sqrt 3-1)y_{\Delta _{a_1}}+\frac{\sqrt 3}{4}(\sqrt 3+1)
    y_{\Delta _{a_2}}\right \} a _1a _{13}
   -\frac{\sqrt 3}{2}y_{\Delta _b}a _5a _{14}\right ]v_d  \nonumber \\
   &\ -3\left [\left \{ \frac{1}{4}(\sqrt 3+1)\bar y_{\Delta _{c_1}}
   -\frac{\sqrt 3}{4}(\sqrt 3-1)\bar y_{\Delta _{c_2}}\right \} a _1a _5
   -\frac{1}{2}\bar y_{\Delta _d}a _{13}a _{14}\right ]v_d , \nonumber \\
\epsilon_{31}&=-y_{\Delta _e}a _5a _9v_d-3\bar y_{\Delta _f}a _9a _{13}v_d , \nonumber \\
\epsilon_{33}&=y_{\Delta _e}a _5a _9v_d .
\label{correction}
\end{align}
Since $\epsilon_{ij}$'s are given as relevant linear combinations of $a_k a_l$'s 
and all $a_i's$ are the same order, 
these are expected to be the same order, assuming Yukawa couplings 
are of order one. Therefore, the magnitude of $\epsilon_{ij}$'s 
are denoted to be ${\cal O}(\tilde a^2v_d)$, 
which is expected to be ${\cal O}(a_i^2v_d)$. 
The charged lepton is diagonalized by 
the left-handed mixing matrix $U_E$ and the right-handed one $V_E$  as
\begin{eqnarray}
V_E^\dagger M_\ell U_E=M_\ell^\text{diag},
\end{eqnarray}
where $M_\ell^\text{diag}$ is a diagonal matrix.
These mixing matrices can be written by
\begin{eqnarray}
\label{VEUE}
\begin{split}
V_E&=
\begin{pmatrix}
       \cos 60^\circ &  \sin 60^\circ &  0\\
    -\sin 60^\circ & \cos 60^\circ  &  0\\
    0& 0& 1
\end{pmatrix}
\times
\begin{pmatrix}
       1 &  \frac{\tilde a^2}{\lambda^2} &  \tilde a\\
    -\frac{\tilde a^2}{\lambda^2}-\tilde a^2  & 1  &  \tilde a\\
    -\tilde a+ \frac{\tilde a^3}{\lambda^2} 
& -\tilde a-  \frac{\tilde a^3}{\lambda^2}& 1
\end{pmatrix},
\\
U_E&=
 \begin{pmatrix}
       1 &  \frac{\tilde a}{\lambda} &  \tilde a\\
    -\frac{\tilde a}{\lambda}-\tilde a^2  & 1  &  \tilde a\\
    -\tilde a+ \frac{\tilde a^2}{\lambda} 
& -\tilde a-  \frac{\tilde a^2}{\lambda}& 1
\end{pmatrix}.
\end{split}
\end{eqnarray}
Taking the next-to-leading order, the electron has non-zero mass, namely
\begin{eqnarray}
m_e^2 &\simeq 
\frac{3}{2}\left (\frac{1}{6}\epsilon _{11}^2
-\frac{1}{\sqrt 3}\epsilon _{11}\epsilon _{21}+\frac{1}{2}\epsilon _{21}^2\right )\simeq {\cal O}(\tilde a^4 v_d^2).
\end{eqnarray}

Next, the down-type quark mass matrix including the next-to-leading order is
\begin{equation}
M_d\simeq
\begin{pmatrix}
\bar\epsilon _{11} & \bar\epsilon _{21} & \bar\epsilon _{31} \\
\frac{\sqrt 3m_s}{ 2}+\bar\epsilon_{12} & \frac{m_s}{2}+\bar\epsilon_{22} 
& \bar\epsilon _{32} \\
\bar\epsilon _{13} & \bar\epsilon _{23} & m_b+\bar\epsilon_{33}
\end{pmatrix},
\label{nextleading}
\end{equation}
where $\bar\epsilon_{ij}$'s are given 
by replacing  $\bar y_{\Delta_i}$  with  
 $-\bar y_{\Delta_i}/3\ (i=c_1,c_2,d,f)$ in Eq.~(\ref{correction}).
Since the alignment $a _1=a _2$ is taken as in  Eq. (\ref{vev}), 
the mass matrix of the up-type quarks is given as 
\begin{equation}
 M_u=v_u
\begin{pmatrix}
2y_{\Delta _{a_1}}^ua _1^2+y_{\Delta _{b}}^ua_{14}^2 & y_{\Delta _{a_2}}^ua _1^2 & y_1^ua _1 \\
y_{\Delta _{a_2}}^ua _1^2 & 2y_{\Delta _{a_1}}^ua _1^2+y_{\Delta _{b}}^ua_{14}^2 & y_1^ua _1 \\
y_1^ua _1 & y_1^ua _1 & y_2^u+y_{\Delta _c}^ua _9^2
\end{pmatrix}.
\end{equation}
Therefore, the CKM matrix $V^0$ at the GUT scale can be written as 
\begin{equation}
V^{0}= U_u^\dagger 
\begin{pmatrix}     
1 & 0 & 0 \\
0 & e^{-i \rho} & 0 \\
0 & 0 & 1
\end{pmatrix}U_d \ ,
\end{equation}
where the left-handed mixing matrix of the up quarks $U_u$ is given as
\begin{eqnarray}
U_u=
\begin{pmatrix}
\cos 45^\circ & \sin 45^\circ & 0 \\
-\sin 45^\circ & \cos 45^\circ & 0 \\
0 & 0 & 1
\end{pmatrix} 
\begin{pmatrix}     
1 & 0 & 0 \\
0 & r_t & r_c \\
0 & -r_c & r_t
\end{pmatrix},
\end{eqnarray}
and the left-handed mixing matrix of the down  quarks $U_d$ is given as
\begin{equation}
U_d=
\begin{pmatrix}
\cos 60^\circ & \sin 60^\circ & 0 \\
-\sin 60^\circ & \cos 60^\circ & 0 \\
0 & 0 & 1
\end{pmatrix} 
\begin{pmatrix}
1 & \theta_{12}^d & \theta_{13}^d \\ 
-\theta _{12}^d-\theta _{13}^d\theta _{23}^d  & 1 & \theta _{23}^d \\
-\theta _{13}^d+\theta _{12}^d\theta _{23}^d  & -\theta _{23}^d-\theta _{12}^d\theta _{13}^d & 1 \\
\end{pmatrix}.
\end{equation}
Here,  $r_c=\sqrt{m_c/(m_c+m_t)}$, $r_t=\sqrt{m_t/(m_c+m_t)}$, 
and the phase $\rho$ is an arbitrary parameter originating from  complex
Yukawa couplings. Magnitudes of $\theta_{ij}^d$ are given as 
\begin{eqnarray}
&&\theta _{12}^d=\mathcal{O}\left (\frac{m_d}{m_s}\right )
=\mathcal{O}\left (0.05\right ),\ \  
\theta _{13}^d=\mathcal{O}\left (\frac{m_d}{m_b}\right )
=\mathcal{O}\left (0.005\right ),\ 
\nonumber\\
&&\theta _{23}^d=\mathcal{O}\left (\frac{m_d}{m_b}\right )
=\mathcal{O}\left (0.005\right ).
\label{dangles}
\end{eqnarray}
At the leading order, the Cabibbo angle is derived as $15^\circ$ 
and it can be naturally fitted to the observed value by including
the next-to-leading  order as follows:
\begin{eqnarray}
V_{us}^{0}
&\simeq \theta _{12}^d\cos 15^\circ +\sin 15^\circ.
\end{eqnarray}

Magnitudes of  $a_i = \langle \chi_i \rangle / \Lambda$ 
are determined by putting  the quark and lepton masses, except
for $a_{14}$, which appears at the next-to-leading order.
These are given as
\begin{align}
&a_ 3=a_8=a_{10}=a_{11}=a_{12}=0, 
\qquad 
a_1=a_2\simeq\sqrt{\frac{m_c}{2\left |y_{\Delta _{a_2}}^u-\frac{{y_1^u}^2}{y_2^u}\right |v_u}}~, 
\nonumber\\
&a_4 = \frac{(y_1^D\lambda )^2(m_3-m_1)m_2M}{6y_2^N{y_2^D}^2m_1m_3\Lambda },
\qquad
a _5 =a_6 =a_7 = \frac{\sqrt{m_2M}}{\sqrt 3y_2^Dv_u}, 
\nonumber \\
& a_9= \frac{m_\mu }{\sqrt 6|\bar{y_1}|\lambda v_d}, 
\qquad a_{13} = \frac{m_\tau }{y_2v_d}\ .
\label{alphas}
\end{align}
where  masses of quarks and leptons are given at the GUT scale.

\section{Slepton mass matrices}

We study SUSY breaking terms
in the framework of $S_4 \times Z_4 \times U(1)_{FN}$
to predict slepton mass matrices.
We consider the gravity mediation within the 
framework of supergravity theory.
We assume that 
non-vanishing $F$-terms of gauge and flavor singlet (moduli) fields $Z$ 
and gauge singlet fields $\chi_i$ $(i=1,\cdots,14)$ 
contribute to the SUSY breaking.
Their $F$-components are written as 
\begin{eqnarray}
F^{\Phi_k}= - e^{ \frac{K}{2M_p^2} } K^{\Phi_k \bar{I} } \left(
  \partial_{\bar{I}} \bar{W} + \frac{K_{\bar{I}}} {M_p^2} \bar{W} \right) ,
\label{eq:F-component}
\end{eqnarray}
where $M_p$ is the Planck mass, $W$ is the superpotential, 
$K$ denotes the K\"ahler potential, $K_{\bar{I}J}$ denotes 
second derivatives by fields, 
i.e. $K_{\bar{I}J}={\partial}_{\bar{I}} \partial_J K$
and $K^{\bar{I}J}$ is its inverse. 
Here the fields ${\Phi_k}$ correspond to the moduli fields $Z$ and 
gauge singlet fields $\chi_i$.
The VEVs of $F_{\Phi_k}/\Phi_k$  are estimated as 
$\langle F_{\Phi_k}/ \Phi_k \rangle = {\cal O}(m_{3/2})$, where
$m_{3/2}$ denotes the gravitino mass, which is obtained as 
$m_{3/2}= \langle e^{K/2M_p^2}W/M_p^2 \rangle$.


First, let us study soft scalar masses.
Within the framework of supergravity theory,
soft scalar mass squared is obtained as~\cite{Kaplunovsky:1993rd}
\begin{eqnarray}
m^2_{\bar{I}J} K_{{\bar{I}J}}= m_{3/2}^2K_{{\bar{I}J}} 
+ |F^{\Phi_k}|^2 \partial_{\Phi_k}  
\partial_{  \bar{\Phi_k} }  K_{\bar{I}J}-
|F^{\Phi_k}|^2 \partial_{\bar{\Phi_k}}  K_{\bar{I}L} \partial_{\Phi_k}  
K_{\bar{M} J} K^{L \bar{M}}.
\label{eq:scalar}
\end{eqnarray}
The invariance under the  $S_4 \times Z_4 \times U(1)_{FN}$
flavor symmetry 
as well as the gauge invariance requires the following form 
of the K\"ahler potential as 
\begin{equation}
K = Z^{(L)}(\Phi )\sum_{i=e,\mu,\tau } |L_i|^2 + 
Z_{(1)}^{(R)}(\Phi )\sum_{i=e,\mu }|R_i|^2 +Z_{(2)}^{(R)}(\Phi )|R_\tau |^2, 
\label{eq:Kahler}
\end{equation}
at the lowest level, where $Z^{(L)}(\Phi)$ and $Z_{(1),(2)}^{(R)}(\Phi)$ are 
arbitrary functions of the singlet fields $\Phi$.
By use of Eq.~(\ref{eq:scalar}) with 
the K\"ahler potential in Eq.~(\ref{eq:Kahler}), 
we obtain the following matrix form 
of soft scalar masses squared for left-handed and 
right-handed charged sleptons,
\begin{eqnarray}
(m_{\tilde L}^2)_{ij} =
\left(
  \begin{array}{ccc}
m_{L}^2   &  0 &  0 \\ 
0   & m_{L}^2  & 0  \\ 
0   &  0  & m_{L}^2   \\ 
\end{array} \right ),
\quad
(m_{\tilde R}^2)_{ij} 
 = 
\left(
  \begin{array}{ccc}
m_{R(1)}^2   &  0 &  0 \\ 
0   &  m_{R(1)}^2 & 0  \\ 
0   & 0   & m_{R(2)}^2   \\ 
\end{array} \right ).
\label{eq:soft-mass-1}
\end{eqnarray}
That is, three left-handed slepton masses are degenerate, and 
two right-handed slepton masses  are degenerate.
These predictions  would be obvious because 
the left-handed sleptons  form  a triplet of  $S_4$, 
and the right-handed sleptons form  a doublet and a singlet of $S_4$.
These  predictions hold exactly before $S_4\times Z_4\times U(1)_{FN}$
is broken, 
but its breaking gives next-to-leading terms in the slepton mass matrices.

Next, we study effects due to  $S_4 \times Z_4\times U(1)_{FN}$  breaking 
by $\chi_i$.
That is, we estimate corrections to the K\"ahler potential 
including  $\chi_i$.
Since each VEV is taken as the same order, 
the breaking scale can be characterized by the 
average of VEVs  $a_i\Lambda$.

Since the right-handed charged leptons $(R_e^c,R_\mu^c)$ 
are assigned to ${\bf 2}$ and 
its conjugate representation is itself ${\bf 2}$. 
Similarly, the left-handed charged leptons 
$(L_e,L_\mu,L_\tau)$ are assigned to ${\bf 3}$ and 
its conjugation is ${\bf 3}$.  Therefore,
for the left-handed sector,  higher dimensional terms are given as
\begin{align}
\Delta K_L 
&= \sum _{i=1,3} Z_{\Delta _{a_i}}^{(L)}(\Phi )
 (L_e,L_\mu ,L_\tau ) \otimes (L_e^c,L_\mu ^c,L_\tau ^c)
\otimes (\chi _i,\chi _{i+1})\otimes (\chi _i^c,\chi _{i+1}^c)/\Lambda ^2
\nonumber \\
&\ +\sum _{i=5,8,11} Z_{\Delta _{b_i}}^{(L)}(\Phi )
 (L_e,L_\mu ,L_\tau ) \otimes (L_e^c,L_\mu ^c,L_\tau ^c)
\otimes (\chi _i,\chi _{i+1},\chi _{i+2})\otimes (\chi _i^c,\chi _{i+1}^c,\chi _{i+2}^c)/\Lambda ^2
\nonumber \\
&\ + Z_{\Delta _c}^{(L)}(\Phi )
 (L_e,L_\mu ,L_\tau ) \otimes (L_e^c,L_\mu ^c,L_\tau ^c)
\otimes \chi _{14}\otimes \chi _{14}^c/\Lambda ^2
\nonumber \\
&\ +Z_{\Delta _d}^{(L)}(\Phi )
 (L_e,L_\mu ,L_\tau ) \otimes (L_e^c,L_\mu ^c,L_\tau ^c)
\otimes \Theta \otimes \Theta ^c/\bar \Lambda ^2.
\end{align}
For example, 
higher dimensional terms 
including $(\chi _1,\chi_2)$ and 
 $(\chi _5, \chi _6 ,\chi_7)$ are explicitly written  as
\begin{align}
\Delta K_L^{\left [\chi _1,\chi _5\right ]} &
= Z_{\Delta _{a_1}}^{(L)}(\Phi )\left [\frac{\sqrt 2|\chi _1|^2}{\Lambda ^2}(|L_\mu |^2-|L_\tau |^2)\right ] \nonumber \\
&\ +Z_{\Delta _{b_5}}^{(L)}(\Phi )\left [\frac{2|\chi _5|^2}{\Lambda ^2}
(L_\mu L_\tau ^\ast +L_\tau L_\mu ^\ast +L_eL_\tau ^\ast +L_\tau L_e^\ast +L_eL_\mu ^\ast +L_\mu L_e^\ast )\right ].
\end{align}

When we take into account   corrections from all $\chi_i \chi_j^*$ 
to the K\"ahler potential, 
the soft scalar masses squared for left-handed charged sleptons 
have the following corrections, 
\begin{equation}
(m_{\tilde L}^2)_{ij}=\begin{pmatrix}
                              m_L^2+\tilde a_{L1} ^2m_{3/2}^2 & k_La_5 ^2m_{3/2}^2 & k_La_5 ^2m_{3/2}^2 \\
                              k_La_5 ^2m_{3/2}^2 & m_L^2+\tilde a_{L2} ^2m_{3/2}^2 & k_La_5 ^2m_{3/2}^2 \\
                              k_La_5 ^2m_{3/2}^2 & k_La_5 ^2m_{3/2}^2 & m_L^2+\tilde a_{L3} ^2m_{3/2}^2
                           \end{pmatrix},
\end{equation}
where $k_L$ is a parameter of order $1$,
and $\tilde a_{Lk}^2(k=1,2,3)$ are linear combinations of $a_i a_j$'s.

For the right-handed sector, 
higher dimensional terms are given as
\begin{align}
\Delta K_R 
&= \sum _{i=1,3} Z_{\Delta _{a_i}}^{(R)}(\Phi )
 (R_e,R_\mu ) \otimes (R_e^c,R_\mu ^c)
\otimes (\chi _i,\chi _{i+1})\otimes (\chi _i^c,\chi _{i+1}^c)/\Lambda ^2
\nonumber \\
&\ +\sum _{i=5,8,11} Z_{\Delta _{b_i}}^{(R)}(\Phi )
 (R_e,R_\mu ) \otimes (R_e^c,R_\mu ^c) 
\otimes (\chi _i,\chi _{i+1},\chi _{i+2})\otimes (\chi _i^c,\chi _{i+1}^c,\chi _{i+2}^c)/\Lambda ^2
\nonumber \\
&\ + Z_{\Delta _c}^{(R)}(\Phi )
 (R_e,R_\mu ) \otimes (R_e^c,R_\mu ^c) 
\otimes \chi _{14}\otimes \chi _{14}^c/\Lambda ^2
\nonumber \\
&\ + Z_{\Delta _d}^{(R)}(\Phi )
 (R_e,R_\mu ) \otimes R_\tau ^c 
\otimes (\chi _1,\chi _2)/\Lambda ^2
+ Z_{\Delta _e}^{(R)}(\Phi )
 (R_e^c,R_\mu ^c ) \otimes R_\tau 
\otimes (\chi _1^c,\chi _2^c)/\Lambda ^2
\nonumber \\
&\ +\sum_{i=1,3} Z_{\Delta _{f_i}}^{(R)}(\Phi )
 R_\tau \otimes R_\tau ^c
\otimes (\chi _i,\chi _{i+1})\otimes (\chi _i^c,\chi _{i+1}^c)/\Lambda ^2
\nonumber \\
&\ +\sum _{i=5,8,11} Z_{\Delta _{g_i}}^{(R)}(\Phi )
 R_\tau  \otimes R_\tau ^c
\otimes (\chi _i,\chi _{i+1},\chi _{i+2})\otimes (\chi _i^c,\chi _{i+1}^c,\chi _{i+2}^c)/\Lambda ^2
\nonumber \\
&\ + Z_{\Delta _h}^{(R)}(\Phi )
 R_\tau  \otimes R_\tau ^c
\otimes \chi_{14}\otimes \chi_{14}^c/\Lambda ^2
+ Z_{\Delta _i}^{(R)}(\Phi )
 (R_e,R_\mu ) \otimes (R_e^c,R_\mu ^c) 
\otimes \Theta \otimes \Theta ^c/\bar \Lambda ^2
\nonumber \\
&\ + Z_{\Delta _j}^{(R)}(\Phi )
 R_\tau  \otimes R_\tau ^c
\otimes \Theta \otimes \Theta ^c/\bar \Lambda ^2.
\end{align}

In the same way, the right-handed charged slepton mass matrix 
can be written as
\begin{equation}
(m_{\tilde R}^2)_{ij}=\begin{pmatrix}
                              m_{R(1)}^2+\tilde a_{R11} ^2m_{3/2}^2 & \tilde a^2_{R12}m_{3/2}^2 & k_Ra_1m_{3/2}^2 \\
                              \tilde a^2_{R12}m_{3/2}^2 & m_{R(1)}^2+\tilde a_{R22} ^2m_{3/2}^2 & k_Ra_1m_{3/2}^2 \\
                              k_R^*a_1m_{3/2}^2 & k_R^*a_1m_{3/2}^2 & m_{R(2)}^2+\tilde a^2_{R33}m_{3/2}^2
                           \end{pmatrix},
\label{RR}
\end{equation}
where $k_R$ is  a parameter  of order one,
and $\tilde a_{Rij}^2$ are linear combinations of $a_i a_j$'s.


 In order to estimate the magnitude of the flavor changing neutral current 
(FCNC), we move to the super-CKM basis
by diagonalizing the charged lepton mass matrix including  next-to-leading 
terms.
 For the left-handed slepton mass matrix, we get as
\begin{eqnarray}
(m_{\tilde L}^2)_{ij}^{(SCKM)}=U^\dagger_E (m_{\tilde L}^2)_{ij} U_E ,
\end{eqnarray}
and for the right-handed slepton mass matrix, we  get as
\begin{eqnarray}
(m_{\tilde R}^2)_{ij}^{(SCKM)}=V^\dagger_E (m_{\tilde R}^2)_{ij} V_E ,
\end{eqnarray}
where the  mixing matrices $V_E$ and $U_E$ are given in  Eqs. (\ref{VEUE}).


Let us study scalar trilinear couplings, i.e. 
the so called A-terms.
The A-terms among left-handed and right-handed sleptons 
and Higgs scalar fields are obtained in the gravity mediation 
as~\cite{Kaplunovsky:1993rd}
\begin{equation}
h_{IJ} {L}_J {R}_I H_K =  \sum_{K={\bar 5},\ 45}
h^{(Y)}_{IJK}{L}_J {R}_I H_K  + h^{(K)}_{IJK}{L}_J {R}_I H_K ,
\label{eq:A-term}
\end{equation}
where 
\begin{eqnarray}
h^{(Y)}_{IJK} &=& F^{\Phi_k} \langle \partial_{\Phi_k} \tilde{y}_{IJK}
\rangle ,  
\nonumber \\
h^{(K)}_{IJK}{L}_J {R}_I H_K &=& - 
\langle \tilde{y}_{LJK} \rangle {L}_J {R}_I H_K F^{\Phi_k} K^{L\bar{L}}
\partial_{\Phi_k} K_{\bar{L}I}  \\
& &  -
\langle \tilde{y}_{IMK} \rangle {L}_J {R}_I H_d F^{\Phi_k} K^{M\bar{M}} 
\partial_{\Phi_k} K_{\bar{M}J}  \nonumber  \\ 
& & -   \langle \tilde{y}_{IJK} \rangle {L}_J {R}_I H_K F^{\Phi_k} K^{H_d}
\partial_{\Phi_k} K_{H_K}, \nonumber  
\label{eq:A-term-2}
\end{eqnarray}
and $K_{H_K}$ denotes the K\"ahler metric of $H_K$.
In addition,  effective Yukawa couplings $\tilde{y}_{IJK}$
are written as
\begin{eqnarray}
\tilde{y}_{IJK}
 = 
-3y_1
\begin{pmatrix}0 & a_9/\sqrt 2 & 0 \\ 
           0 & a_9/\sqrt 6 & 0 \\
                 0  & 0 & 0   \\
 \end{pmatrix} 
+y_2
\begin{pmatrix} 0 & 0 & 0 \\ 
               0 & 0 & 0 \\
                 0 & 0 & a_{13} \\
 \end{pmatrix}, 
\label{ME}
 \end{eqnarray}
then we have
\begin{eqnarray}
h^{(Y)}_{IJK}
 = 
-\frac{3y_1}{\Lambda }
\begin{pmatrix}0 & \tilde F^{a _9}/\sqrt 2 & 0 \\ 
           0 & \tilde F^{a _9}/\sqrt 6 & 0  \\
                 0  & 0 & 0   \\
 \end{pmatrix}
+\frac{y_2}{\Lambda }
\begin{pmatrix} 0 & 0 & 0 \\ 
               0 & 0 & 0 \\
                0 & 0 & \tilde F^{a _{13}} \\
 \end{pmatrix},
\label{ME}
\end{eqnarray}
where $\tilde F^{a_i}=F^{a_i}/a_i$ and 
$\tilde F^{a_i}/\Lambda = {\cal O}(m_{3/2})$.

 By use of the lowest level of the  K\"ahler potential,
 we estimate $h^{(K)}_{IJK}$ as
\begin{equation}
h^{(K)}_{IJK} = \tilde y_{IJK} (A^R_I+A^L_J),
\end{equation}
where  
 we estimate $A^L_1=A^L_2=A^L_3 
=F^{\tilde a_i}/(a_i\Lambda) \simeq \mathcal{O}(m_{3/2})$. 
The magnitudes of $A^R_1= A^R_2$ and $A^R_3$ are also $\mathcal{O}(m_{3/2})$.
Furthermore, we should take into account next-to-leading terms of 
the K\"ahler potential including $\chi_i$.
These correction terms appear all entries so that their  magnitudes
 are suppressed in ${\cal O}(\tilde a)$
compared with the leading term.  Then, we obtain 
\begin{equation}
 (m_{LR}^2)_{ij} \simeq m_{3/2}
\begin{pmatrix}
\tilde a_{LR11} ^2v_d & c_1\frac{\sqrt{3}m_\mu}{2} & \tilde a_{LR13} ^2v_d \\ 
\tilde a_{LR21} ^2v_d & c_1\frac{m_\mu}{2} & \tilde a_{LR23} ^2v_d \\
\tilde a_{LR31} ^2v_d & \tilde a_{LR32} ^2v_d & c_2m_\tau 
\end{pmatrix},
\end{equation}
where  $\tilde a_{LRij}^2$ are linear combinations of $a_i a_j$'s,
and $c_1$ and $c_2$ are of order one parameters.
Moving to the super-CKM basis, 
we have
\begin{align}
 (m_{LR}^2)^{SCKM}_{ij} =U_E^\dagger  (m_{LR}^2)_{ij} V_E
 \simeq m_{3/2}
\begin{pmatrix}
\mathcal{O} \left (\tilde a^2v_d\right ) 
& \mathcal{O} \left (\tilde a^2v_d\right ) 
& \mathcal{O} \left (\tilde a^2v_d\right ) \\ 
\mathcal{O} \left (\tilde a^2v_d\right ) 
& \mathcal{O}(m_\mu) & \mathcal{O} \left(\tilde a ^2v_d\right ) \\
\mathcal{O} \left (\tilde a^2v_d\right ) 
& \mathcal{O} \left (\tilde a^2v_d\right ) & \mathcal{O}  (m_\tau)
\end{pmatrix}.
\end{align}


\section{Renormalization group effect}

In this section, we consider the running effects of 
slepton mass matrices, A-terms, and Yukawa couplings
 from the GUT scale $m_{\text{GUT}}$ down to the electroweak scale $m_W$. 
The renormalization group (RG) equations are given by 
\cite{Martin:1993zk,Hisano:1995cp};
\begin{eqnarray}
\begin{split}
16\pi^2 \frac{d}{d t} \left( {m}^2_{L} \right)_{ij}
=&   -\left( \frac{6}{5} g_1^2 \left| M_1 \right|^2
+ 6 g_2^2 \left| M_2 \right|^2 \right) \delta_{ij}
-\frac{3}{5} g_1^2~S~\delta_{ij}
\\
&+  \left (( {m}^2_{L} ) {Y}_e^{\dagger} {Y}_e
+ {Y}_e^{\dagger} {Y}_e ( {m}^2_{L} ) \right)_{ij} 
\\
&+ 2 \left( {Y}_e^{\dagger} ({m}^2_{R} ) {Y}_e
           +{m}^2_{H_d} {Y}_e^{\dagger} {Y}_e
+{A}_e^{\dagger} {A}_e \right)_{ij} \ ,
\\
16\pi^2 \frac{d}{d t} \left( {m}^2_{R} \right)_{ij} =&
- \frac{24}{5} g_1^2 \left| M_1 \right|^2 \delta_{ij} + \frac{6}{5} g_1^2~S~\delta_{ij} 
\\
&+ 2 \left ( ( {m}^2_{R} )  {Y}_e {Y}_e^{\dagger} 
+ {Y}_e {Y}_e^{\dagger} ( {m}^2_{R} ) \right)_{ij}
\\
\label{RGE}
&+ 4 \left( {Y}_e ( {m}^2_{L} ) {Y}_e^{\dagger} + {m}^2_{H_d}
{Y}_e {Y}_e^{\dagger} +  {A}_e {A}_e^{\dagger} \right)_{ij}~, 
\\
16\pi^2 \frac{d}{d t}  \left( {A}_e \right)_{ij} =&
 \left( -\frac{9}{5} g_1^2 -3 g_2^2
+ 3 {\rm Tr} ( {Y}_d^{\dagger} {Y}_d )
+   {\rm Tr} ( {Y}_e^{\dagger} {Y}_e ) \right )  \left({A}_{e}\right)_{ij} 
\\
&+ 2 \left(
\frac{9}{5} g_1^2 M_1 + 3 g_2^2 M_2
+ 3 {\rm Tr} ( {Y}_d^{\dagger} {A}_d)
+   {\rm Tr} ( {Y}_e^{\dagger} {A}_e) \right) {Y}_{e_{ij}} 
\\
&+ 4 \left( {Y}_e {Y}_e^{\dagger} {A}_e \right)_{ij}
+ 5 \left({A}_e {Y}_e^{\dagger} {Y}_e \right)_{ij}~, 
\\
16\pi^2 \frac{d}{d t} {Y}_{e_{ij}} =& \left ( -\frac{9}{5} g_1^2 - 3 g_2^2
  + 3 \,{\rm Tr} ( {Y}_d {Y}_d^{\dagger})
  +   {\rm Tr} ( {Y}_e {Y}_e^{\dagger})
\right ) {Y}_{e_{ij}}
+ 3 \, \left( {Y}_e {Y}_e^{\dagger} {Y}_e \right)_{ij}\ .
\end{split}
\label{RGE}
\end{eqnarray}
In these expressions,  $g_{1,2}$ are the gauge couplings of 
SU(2)$_L\times U(1)_Y$, $t=\ln\mu/\mu_0$, 
$M_{1,2}$ are the corresponding gaugino mass terms, 
${Y}_{e,d}\equiv M_{l,d}/v_d$ 
are the Yukawa couplings for charged leptons and down quarks, 
${A}_e= ({m}^2_{LR})/v_d$,  and
\begin{equation}
S = {\rm Tr} ({m}^2_{qL} + {m}^2_{dR}- 2 {m}^2_{uR}
- {m}^2_{L} + {m}^2_{R} ) - {m}^2_{H_d}
+ {m}^2_{H_u} \nonumber,
\end{equation}
where ${m}^2_{qL}$, ${m}^2_{dL}$, ${m}^2_{uR}$ are mass matrices of 
squarks and $m_{H_u}$ and $m_{H_d}$  are the Higgs  masses. 
Numerically, the largest contributions of the effect for off diagonal elements of A-term are 
those of gauge couplings. Then we can estimate the running effects 
by 
\begin{eqnarray}
{A}_{e_{ij}} (m_Z)
=\exp\left[ \frac{-1}{16\pi^2}\int_{m_Z}^{m_\text{GUT}} dt
~ \left ( \frac95 g_1^2+3g_2^2 \right )\right ]{A}_{e_{ij}} (m_\text{GUT})
\approx 1.5\times {A}_{e_{ij}} (m_\text{GUT}).
\nonumber\\
\end{eqnarray}

In the SUGRA framework, 
soft masses for all scalar particles have the common scale denoted 
by $m_{\text{SUSY}}$,  and gauginos also have the common scale $m_{1/2}$. 
Therefore, at the GUT scale, we take 
\begin{eqnarray}
M_1 (m_\text{GUT}) = M_2 (m_\text{GUT}) = m_{1/2} \; .
\end{eqnarray}
Effects of RG running lead at the scale $m_W$ 
to  following masses for gauginos
\begin{eqnarray}
\label{gaugino}
M_1(m_W)\simeq\dfrac{\alpha_1(m_W)}{\alpha_1(m_\text{GUT})}M_1(m_\text{GUT}),
\quad
M_2(m_W)\simeq\dfrac{\alpha_2(m_W)}{\alpha_2(m_\text{GUT})}M_2(m_\text{GUT}),
\end{eqnarray}
where $\alpha_i=g_i^2/4\pi$ ($i=1,2$) 
and according to the gauge coupling unification at 
$m_\text{GUT}$, 
$\alpha_1(m_\text{GUT})=\alpha_2(m_\text{GUT})\simeq 1/25$.
Taking into account the RG effect on the average mass scale in $m_L^2$
 and $m_R^2$, we have
\begin{eqnarray}
\label{smass}
m_L^2(m_W)&\simeq& m_L^2(m_\text{GUT})+0.5M_2^2(m_\text{GUT})
+0.04M_1^2(m_\text{GUT}) \simeq m_{\text{SUSY}}^2 +0.54 m_{1/2}^2 ,
\nonumber\\
m_R^2(m_W)&\simeq& m_R^2(m_\text{GUT})+0.15M_1^2(m_\text{GUT}) 
\simeq m_{\text{SUSY}}^2 +0.15 m_{1/2}^2 \; .
\end{eqnarray}
The parameter $\mu$ is given through the requirement of the correct 
electroweak symmetry breaking.
 At the electroweak scale, we have \cite{Feruglio}
\begin{eqnarray}
|\mu|^2\simeq-\dfrac{m_Z^2}{2}+m_{\text{SUSY}}^2\dfrac{1+0.5\tan^2\beta}{\tan^2\beta-1}+m_{1/2}^2\dfrac{0.5+3.5 \tan^2\beta}{\tan^2\beta-1}\;,
\label{defmuSUGRA}
\end{eqnarray}
which is determined by $m_{\text{SUSY}}$, $m_{1/2}$  and $\tan\beta$.



Let us discuss the allowed $\tan\beta$ focusing on the $m_b/m_\tau$ ratio.
For Yukawa couplings, the $b-\tau$ unification is realized 
at the leading order in our model. 
However, the $b-\tau$ unification is deviated when we include
 the next-to-leading order mass operators  due to terms including $H_{45}$ 
as seen in Eq. (\ref{nextsusy}). 
Source terms to cause the deviation for $(3,3)$ element 
can be estimated as $y_{\Delta_e}a_5a_9v_d$ for $\tau$ and 
$-y_{\Delta_e}a_5a_9v_d/3$ for the bottom quark. 
Those become non-negligible compared to the leading term $y_2a_{13}v_d$. 
 Concludingly, the $b-\tau$ unification could be deviated in  several percent.
 
We performed a numerical analysis, supposing that 
the next-to-leading order makes up to $10\%$ deviation of the
$b-\tau$ unification, to find the correct ratio of $m_b/m_\tau$ 
by using the RG equations in Eq. (\ref{RGE})
with the SUSY threshold corrections, 
where top quark mass and the heaviest neutrino masse are chosen to be 
consistent with observed values: $m_t(m_z)=181\pm 13$GeV 
and
$m_{\nu_3}=\sqrt{\Delta m^2_{\rm atm}}$  
\cite{Fusaoka:1998vc,Nakamura:2010zzi}.
As seen in Figure 1 (a), we obtain the correct  $m_b/m_\tau$ ratio if
 $\tan\beta$ is larger than two.
In order to keep small $a_9$ and $a_{13}$, 
the low $\tan\beta$ is preferred. 
We take experimental values of $m_b(m_Z)=3.0\pm0.2$GeV and 
$m_\tau(m_Z)=1.75$GeV \cite{Fusaoka:1998vc,Nakamura:2010zzi}, 
which give $m_b/m_\tau=1.60$--1.83. 
We show the   $m_b/m_\tau$  ration versus  $|y_2|a_{13}$
 in a typical case of $\tan\beta=4.5-5.5$ in Figure 1 (b).
The Yukawa coupling of the bottom quark  
at the GUT scale $|y_2|a_{13}$ is obtained to be $0.03-0.04$. 
In this work, we will calculate  LFV and EDM  for the fixed value 
of  $\tan\beta=5$ in latter sections since a lower  $\tan\beta$ value 
becomes inconsistent with the experiments \cite{Komine}.

\begin{figure}[tb]
\subfigure[]{\includegraphics[width=7cm]{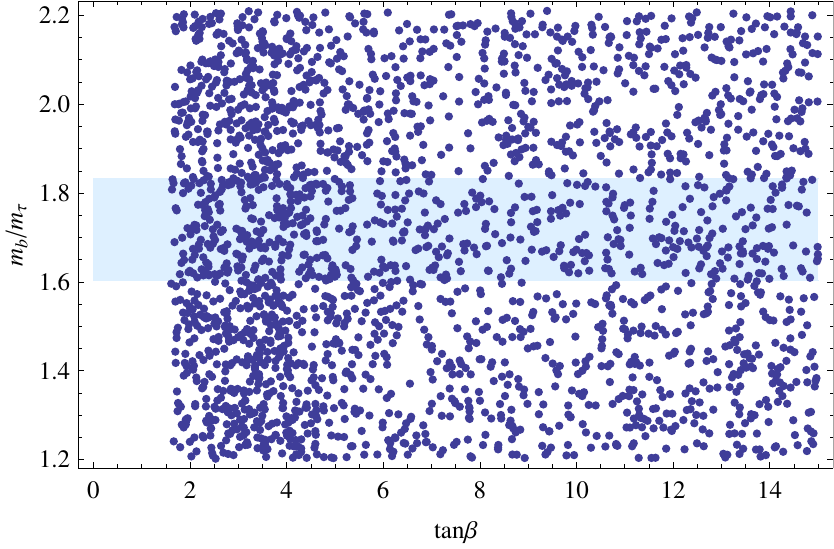}}
\qquad
\subfigure[]{\includegraphics[width=7cm]{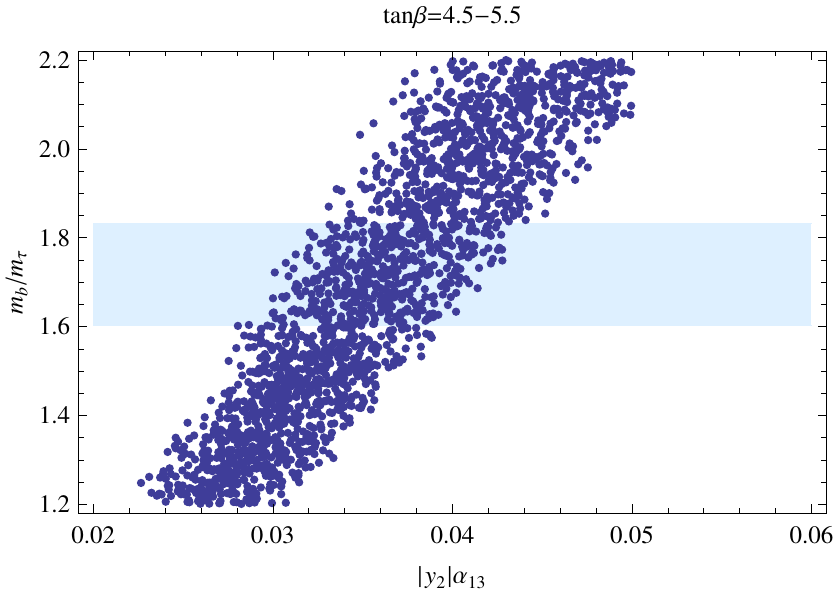}}
\caption{The mass ratio of bottom to tau in the electroweak scale 
is shown versus  (a) $\tan\beta$, and 
(b) $|y_2|a_{13}$ at the GUT scale.  
The shaded region describes the experimentally arrowed region. 
All other parameters such as the top mass and 
neutrino masses are chosen to be consistent with experiments. }
\end{figure}

Now we can estimate  values of $a_i$ by using Eq. (\ref{alphas}).
Putting typical values of quark masses at GUT scale~\cite{Fusaoka:1998vc},
$M=10^{12}~\ \text{GeV}$, $\lambda=0.1$, 
and $\tan\beta=5$ ($v_d\simeq 34~\text{GeV}$, $v_u\simeq 170~\text{GeV}$),
 we have 
\begin{align}
a _1 \sim 3\times 10^{-2},
\quad 
a _4 \sim 10^{-2},
\quad  
a _5 \sim 10^{-2},
\qquad 
a _9 \sim 5\times 10^{-3},
\quad 
a _{13}\sim 3\times 10^{-2},
\label{valuealpha}
\end{align}
where all  Yukawa couplings are assumed to be one.
If we use smaller Yukawa couplings than $1$, these values  of $a_i$
are changed in a factor.
Therefore, the magnitudes of all $a_i$ are supposed to be  order $10^{-2}$.

\section{LFV and EDM in SUSY flavor}

We discuss SUSY flavor phenomena for the lepton sector in the $S_4$ model. 
Mass insertion parameters,  $\delta_\ell^{LL}$, $\delta_\ell^{LR}$,
  $\delta_\ell^{RL}$ and   $\delta_e^{RR}$  are defined by 
\begin{eqnarray}
m_{\tilde\ell}^2
 \begin{pmatrix} \delta_\ell^{LL} & \delta_\ell^{LR} \\ 
                   \delta_\ell^{RL}    & \delta_e^{RR}  \\
 \end{pmatrix}
=
\begin{pmatrix} m_{\tilde L}^2 & m_{LR}^2 \\ 
                   m_{RL}^2    & m_{\tilde R}^2  \\
 \end{pmatrix}
- \text{diag}(m_{\tilde\ell}^2) \ ,
\end{eqnarray}
where $m_{\tilde\ell}$ is an average slepton mass. 
In the SCKM basis, they are estimated as
\begin{eqnarray}
&&\delta_\ell^{LL}
=\begin{pmatrix}
                              \mathcal{O}(\tilde a ^2) & \mathcal{O}(\tilde a ^2) & \mathcal{O}(\tilde a ^2) \\
                              \mathcal{O}(\tilde a ^2) & \mathcal{O}(\tilde a ^2) & \mathcal{O}(\tilde a ^2) \\
                              \mathcal{O}(\tilde a ^2) & \mathcal{O}(\tilde a ^2) & \mathcal{O}(\tilde a ^2)
                           \end{pmatrix},
\quad
\delta_e^{RR}
=\begin{pmatrix}
                              \mathcal{O}(\tilde a ^2) & \mathcal{O}(\tilde a ^2) & \mathcal{O}(a_1) \\
                              \mathcal{O}(\tilde a ^2) & \mathcal{O}(\tilde a ^2) & \mathcal{O}(a_1) \\
                              \mathcal{O}(a_1) & \mathcal{O}(a_1) & \mathcal{O}(\tilde a ^2)
                           \end{pmatrix},
\nonumber\\
&&\delta_\ell^{LR}
=\frac{1}{m_{\tilde\ell}}
\begin{pmatrix}
                              \mathcal{O}(\tilde a ^2(1+\frac{\mu\tan\beta}{m_{\tilde\ell}})v_d) & \mathcal{O}(\tilde a ^2v_d) & \mathcal{O}(\tilde a ^2v_d) \\
                              \mathcal{O}(\tilde a ^2v_d) & \mathcal{O}(m_\mu(1+\frac{\mu\tan\beta}{m_{\tilde\ell}})) & \mathcal{O}(\tilde a ^2v_d) \\
                              \mathcal{O}(\tilde a ^2v_d) & \mathcal{O}(\tilde a ^2v_d) & \mathcal{O}(m_\tau(1+\frac{\mu\tan\beta}{m_{\tilde\ell}}))
                           \end{pmatrix}.
\nonumber\\
\end{eqnarray}
With these parameters, we calculate 
 $\ell_i \rightarrow\ell_j \gamma$ ratios and EDM's of leptons.


In general, when there are 
right-handed neutrinos which couple to the left-handed
neutrinos via Yukawa coupling, the effects from RG
running can also induce off-diagonal elements in the slepton
mass matrix. We have already estimated this effect in the previous work
\cite{Ishimori:2010xk} as 
\begin{equation}
(\delta_\ell^{LL})_{12}=
\frac{6m_0^2}{16\pi ^2m_\text{SUSY}^2}(Y_D^\dagger Y_D)_{12}
\ln \frac{\Lambda }{M}
\simeq \frac{3}{8\pi^2}{y_2^D}^2 a  _5^2 \ln \frac{\Lambda }{M}
\simeq 6\times 10^{-5}\ ,
\end{equation}
where we put 
$m_0=m_\text{SUSY}$, $ a  _5=10^{-2}$, 
$\Lambda =10^{16}$ GeV,  $M=10^{12}$ GeV.
Since the key ingredient  $(Y_D^\dagger Y_D)_{12}$ is rather small such as
  $(Y_D^\dagger Y_D)_{12}={y_2^D}^2 a  _5^2$, the branching ratio
 is suppressed.
 It is concluded that the contribution 
 on  $\mu \rightarrow e\gamma $ from the neutrino sector is 
much smaller than the experimental bound  
$(\delta_\ell^{LL})^{\rm exp}_{12} \leq {\cal O}(10^{-3})$~\cite{Gabbiani:1996hi}.
 Therefore, we neglect the effect of the Dirac neutrinos 
in the following calculations.

\subsection{\boldmath $\mu\to e\gamma$, $\tau\to e\gamma$ 
and $\tau\to \mu\gamma$}

In the framework of SUSY, LFV effects  originate 
from  misalignment between fermion and sfermion mass eigenstates.
 Once non-vanishing off diagonal elements of the slepton mass matrices
are generated in the super-CKM basis,
LFV rare decays like $\ell_i\to\ell_j\gamma$ are naturally induced by one-loop diagrams with the exchange of gauginos 
and sleptons. The present bounds on these processes are summarized 
in Table \ref{tab:lfvtable} \cite{Nakamura:2010zzi}.

\begin{table}[t]
\addtolength{\arraycolsep}{3pt}
\renewcommand{\arraystretch}{1.3}
\centering
\begin{tabular}{|l|l|l|l|}
\hline
Process & BR($\mu \to e\,\gamma$) &  BR($\tau \to e\,\gamma$) &BR($\tau \to \mu\,\gamma$) \\
\hline
Experimental limit 
 & $1.2~ \times~ 10^{-11}$ & $1.1~ \times~ 10^{-7}$ & $6.8~ \times~ 10^{-8}$\\
\hline
\end{tabular}
\caption{\small
Present limits on the lepton flavor violation for each process 
\cite{Nakamura:2010zzi}.}
\label{tab:lfvtable}
\end{table}

The decay $\ell_i\to\ell_j\gamma$ is described by the dipole operator and the corresponding amplitude reads
\cite{Gabbiani:1996hi,Hisano:1995cp,Borzumati:1986qx,Hisano:1995nq}
\begin{eqnarray}
T=m_{\ell_i}\epsilon^{\lambda}\overline{u}_j(p-q)[iq^\nu\sigma_{\lambda\nu}
(A_{L}P_{L}+A_{R}P_{R})]u_i(p)\,,
\end{eqnarray}
where $p$ and $q$ are momenta of the initial lepton $\ell_i$ and of the photon, respectively, 
and $A_{L,R}$ are the two possible amplitudes in this  process. 
The branching ratio of $\ell_{i}\rightarrow \ell_{j}\gamma$ can be written 
as follows:
\begin{eqnarray}
\frac{{\rm BR}(\ell_{i}\rightarrow  \ell_{j}\gamma)}{{\rm BR}(\ell_{i}\rightarrow 
\ell_{j}\nu_i\bar{\nu_j})} =
\frac{48\pi^{3}\alpha}{G_{F}^{2}}(|A_L^{ij}|^2+|A_R^{ij}|^2)\,.
\nonumber
\end{eqnarray}
In the mass insertion approximation, it is found that~\cite{Altmannshofer}
\begin{eqnarray}
\label{MIamplL}
A^{ij}_L
&\simeq&\frac{\alpha_2}{4\pi}
\frac{\left(\delta^{LL}_{\ell}\right)_{ij}}{m_{\tilde \ell}^{2}}\tan{\beta}
~\bigg[
\frac{\mu M_{2}}{(M_{2}^2-\mu^2)}\bigg(f_{2n}(x_2,x_\mu)+f_{2c}(x_2,x_\mu)\bigg)
\nonumber\\
&&+ \tan^2\theta_{W}\,
\mu M_{1}\bigg(\frac{f_{3n}(x_1)}{m_{\tilde \ell}^{2}}+
\frac{f_{2n}(x_1,x_\mu)}{(\mu^2 - M_{1}^2)}\bigg)
\bigg]
+ \frac{\alpha_1}{4\pi}~\frac{\left(\delta^{RL}_{\ell}\right)_{ij}}{m_{\tilde \ell}^2}~
\left(\frac{M_1}{m_{\ell_i}}\right)~2~f_{2n}(x_1)~,
\nonumber\\
A^{ij}_R
&\simeq&
\frac{\alpha_{1}}{4\pi}
\left[
\frac{\left(\delta^{RR}_{e}\right)_{ij}}{m_{\tilde \ell}^{2}}\mu M_{1}\tan{\beta}
\left(\frac{f_{3n}(x_1)}{m_{\tilde \ell}^{2}}-\frac{2f_{2n}(x_1,x_{\mu})}{(\mu^2-M_{1}^2)}\right)
+2\frac{\left(\delta^{LR}_{e}\right)_{ij}}{m_{\tilde\ell}^{2}}~
\left(\frac{M_1}{m_{\ell_i}}\right)~f_{2n}(x_1)
\right]~,
\nonumber\\
\label{form}
\end{eqnarray}
where $\theta_W$ is the weak mixing angle, 
$x_{1,2}=M_{1,2}^2/m_{\tilde \ell}^2$, $x_\mu=\mu^2/m_{\tilde \ell}^2$ 
and $f_{i(c,n)}(x,y)=f_{i(c,n)}(x)-f_{i(c,n)}(y)$. The loop functions 
$f_i$'s are given explicitly as follows:
 \begin{eqnarray}
\begin{split}
f_{2n}(x) &= \frac{-5x^2+4x+1+2x(x+2)\log x}{4(1-x)^4}~, 
\\
f_{3n}(x) &= \frac{1+9x-9x^2-x^3+6x(x+1)\log x}{3(1-x)^5}~,
\\
f_{2c}(x) &= \frac{-x^2-4x+5+2(2x+1)\log x}{2(1-x)^4}~.
\end{split}
\end{eqnarray}

\begin{figure}[tb]
\hspace{-1cm}
\subfigure[]{\includegraphics[width=7cm]{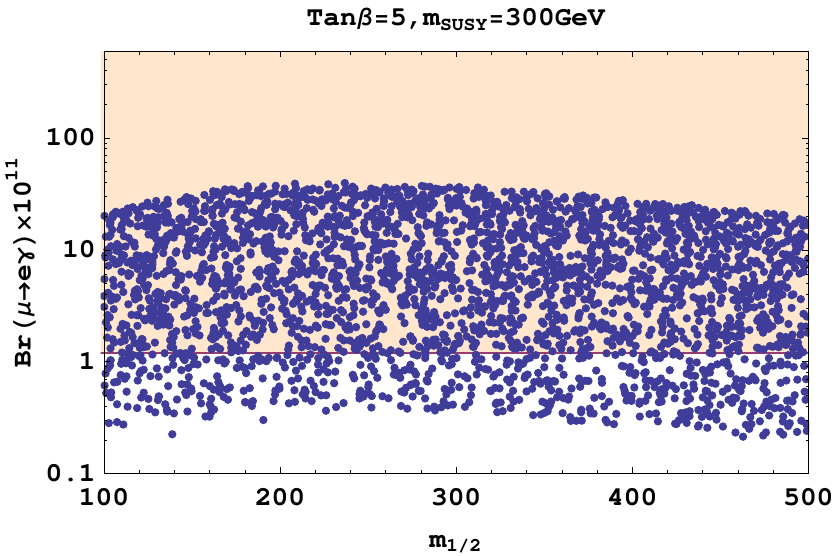}}
\qquad
\subfigure[]{\includegraphics[width=7cm]{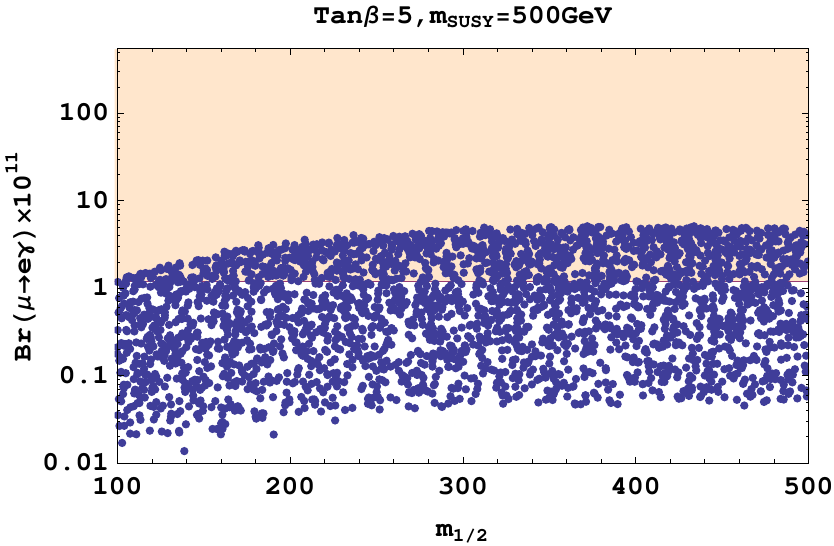}}
\caption{Branching ratio of $\mu\rightarrow e\gamma$ versus
 the gaugino mass parameter $m_{1/2}$ 
for (a) $\tan\beta=5$, $m_\text{SUSY}=300$GeV, and 
 (b) $\tan\beta=5$, $m_\text{SUSY}=500$GeV. 
Shaded regions show exclusion from current experiments, i.e. 
Br$(\mu\rightarrow e\gamma)>1.2\times 10^{-11}$.}
\end{figure}

\begin{figure}[tb]
\subfigure[]{\includegraphics[width=7cm]{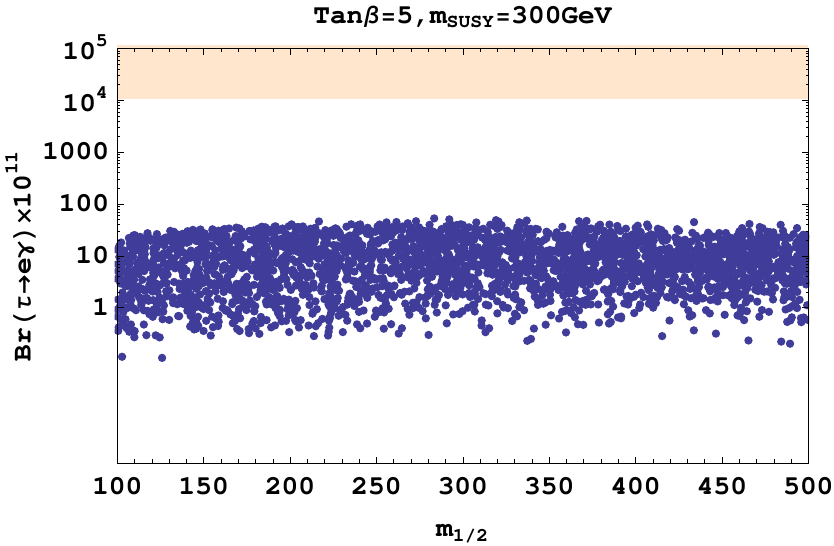}}
\qquad
\subfigure[]{\includegraphics[width=7cm]{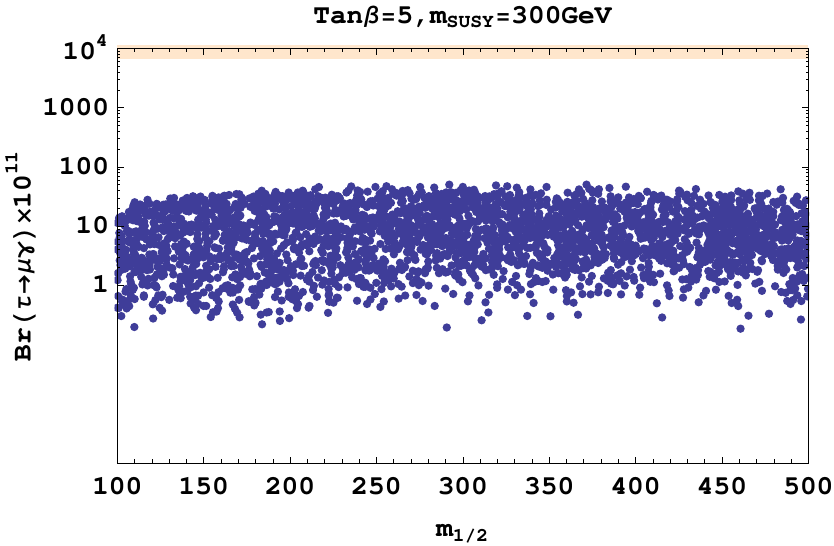}}
\caption{Scattering plots for the branching ratios of 
$\tau\rightarrow e\gamma$ and $\tau\rightarrow \mu\gamma$ 
versus gaugino mass $m_{1/2}$. Both are calculated 
with $\tan\beta=5$ and $m_\text{SUSY}=300$GeV. 
Experimental limits for each processes are $1.1\times 10^{-7}$ and 
$6.8\times 10^{-8}$, respectively. }
\end{figure}

In numerical calculations of the  $\mu\rightarrow e\gamma$ ratio, 
we take $\tan\beta=5$, $m_\text{GUT}=2\times10^{16}$GeV, 
and the SUSY mass scale, $m_{\text{SUSY}}$,
 as $300$GeV or $500$GeV. 
We see the dependence of $m_{1/2}$ up to $500$GeV. 
Gaugino and slepton masses at the electroweak scale 
can be calculated by $m_{1/2}$ and $m_{\text{SUSY}}$ as 
in Eqs. (\ref{gaugino}) and (\ref{smass}). 
Similarly, $\mu$ parameter is also calculable by putting $\tan\beta$, 
$m_{1/2}$, and $m_{\text{SUSY}}$, see Eq. (\ref{defmuSUGRA}). 
We vary absolute values of Yukawa couplings
from $0.1$ to $1$ in the calculation. 
Then, we obtain the numerical result of the branching ratio which 
is illustrated in Figures  2 (a) and (b). 
In the  branching ratio of Eq. (\ref{form}), 
there are terms which proportional to 
the $\mu$-parameter, which increases as the gaugino mass $m_{1/2}$ increases  as seen in Eq. (\ref{defmuSUGRA}). 
Therefore, our predicted branching  ratio does not necessarily decrease 
as $m_{1/2}$ increases.  

The predicted  region of the ratio with $m_{\text{SUSY}}=300$GeV, $500$GeV
 lies within  the region of  expected sensitivity at the MEG experiment
 \cite{Adam:2009ci},  concretely, 
$\mathcal{O}(10^{-13})$--$\mathcal{O}(10^{-14})$. 
When $m_{\text{SUSY}}=300$GeV, the branching ratio 
cannot be smaller than $10^{-12}$. 
Increasing  $m_{\text{SUSY}}$  to 500GeV, 
the lowest value of the ratio is about $10^{-13}$.
 Thus we expect the observation of the $\mu\rightarrow e\gamma$ process 
 at the MEG experiment \cite{Adam:2009ci}.

In the same method, we also calculated the branching ratios of 
$\tau\rightarrow e\gamma$ and $\tau\rightarrow \mu\gamma$ as shown 
in Figures 3 (a) and (b). 
All of $\ell_i\rightarrow \ell_j\gamma$ ratios have the same order
due to structures of the slepton mass matrices. 
Predicted ratios of $\tau\rightarrow e\gamma$ and 
$\tau\rightarrow \mu\gamma$ are much  below the current experimental 
bounds. Future experiments such as SuperB cannot 
reach the expected ratios in  our flavor model.

\subsection{Electric dipole moment}

The mass insertion parameters also contribute to the electron EDM 
through one-loop exchange of binos/sleptons. 
The corresponding EDM is given as
\cite{Hisano:2007cz,Hisano:2008hn,Altmannshofer}
\begin{eqnarray}
\label{Eq:lEDM_LO}
\frac{d_{e}}{e}
\!\!=\!\!
-\frac{\alpha_1}{4\pi}\frac{M_1}{m^{2}_{\tilde\ell}}\!\!\!
&\bigg\{&
\!\!\!{\rm Im}
[(\delta^{LR}_{\ell})_{1k}(\delta^{RR}_{e})_{k1} + (\delta^{LL}_{\ell})_{1k}(\delta^{LR}_{\ell})_{k1}]
\,f_{3n}(x_1)
+{\rm Im}[(\delta^{LL}_{\ell})_{1k}(\delta^{LR}_{\ell})_{kl}(\delta^{RR}_{e})_{l1}
\nonumber\\
&+&
(\delta^{LR}_{\ell})_{1k}(\delta^{RR}_{e})_{kl}(\delta^{RR}_{e})_{l1}+
(\delta^{LL}_{\ell})_{1k}(\delta^{LL}_{\ell})_{kl}(\delta^{LR}_{\ell})_{l1}]
\,f_{4n}(x_1)
\bigg\}\,,
\end{eqnarray}
where $k,l=2,3$, $(\delta^{LR}_{\ell})_{33}= -m_{\tau}(A_{\tau}+\mu\tan\beta)/m^{2}_{\tilde\ell}$, and  the loop function $f_{4n}$
 is given as 
\begin{eqnarray}
\begin{split}
f_{4n}(x) &=   \frac{-3-44x+36x^2+12x^3-x^4-12x(3x+2)\log x}{6(1-x)^6}\, .
\end{split}
\end{eqnarray}

Since components $(i,3)$ and $(3,i)$ of $\delta_e^{RR}$ are 
much larger compared to others, 
dominant terms are given as
\begin{eqnarray}
\label{Eq:lEDM_LO}
\frac{d_{e}}{e}
\approx
-\frac{\alpha_1}{4\pi}\frac{M_1}{m^{2}_{\tilde\ell}}
&\bigg\{&
\mathcal{O}(\frac{m_e}{m_{\tilde\ell}}a_1)
\,f_{3n}(x_1)
+\mathcal{O}(\frac{m_\tau}{m_{\tilde\ell}}(1+\frac{\mu\tan\beta}{m_{\tilde\ell}})
a_1\tilde a^2)
\,f_{4n}(x_1)
\bigg\}.
\end{eqnarray}

In the same parameter regions 
 for the calculation of $\ell_i\rightarrow \ell_j \gamma$ ratios, 
we numerically estimate  EDM of leptons. 
We present the result of $|d_e|$  in Figure 4 (a), in which 
 $\tan\beta=5$,
 $m_{\text{SUSY}}=300$GeV and $m_{1/2}=100-500$GeV. 
Since phases of Yukawa coupling constants are important 
in this estimate,
 we randomly choose $0$ to $2\pi$ for phases of all Yukawa couplings. 
The current experimental bound is $1.6\times 10^{-27}e\text{cm}$ 
\cite{Regan:2002ta}, which 
is denoted by shaded region. Without tuning  phase parameters 
our prediction is  below the present experimental bound. 
We expect the observation of the electron EDM 
in the future experiment, in which  the experimental sensitivity will be 
improved as $10^{-31}e\text{cm}$ \cite{DeMille}.

In Figure 5, we show our  predicted region on  
$|d_e|$ and Br$(\mu\rightarrow e\gamma)$ plane for the case 
$m_\text{SUSY}=300$GeV and $m_{1/2}=100$--$300$GeV.  
As one can see from Figure 5,
 our predicted  region of the electron EDM is not so restricted
even if the branching ratio of $\mu \rightarrow e\gamma$ is fixed.
For example, when $\mu \rightarrow e\gamma$ decay
will be observed just below the present experimental bound,
the predicted electron EDM can be large 
 ${\cal O}(10^{-28})e\text{cm}$ or  small 
${\cal O}(10^{-31})e\text{cm}$, compared to the current experimental bound.

We  have also  calculated EDM's of muon and tau. 
Since the components $(i,3)$ and $(3,i)$ of $\delta_e^{RR}$ 
also dominate  EDM's, predictions are not so different from 
$|d_e|$. 
Although there is no  exact relations 
among $|d_e|$, $|d_\mu|$ and $|d_\tau|$ due to different 
Yukawa  couplings, we can say that the magnitudes of them 
are the same order. Numerically, the results of $|d_\mu|$ is 
shown in Figure  6 (a) and $|d_\tau|$ in Figure 6 (b). 

In our calculations of LFV and EDM of leptons,  we have used SUSY parameters
$m_{\rm SUSY}=300,~500$GeV and $m_{\rm 1/2}=100-500$GeV.
In these parameter regions, we have estimated
the SUSY contribution on the anomalous magnetic moment 
 $a_\mu=(g-2)_\mu/2$, in which  the experimental allowed value: 
$\Delta a_\mu=a_\mu^\text{exp}-a_\mu^\text{SM}
\simeq (3\pm 1)\times 10^{-9}$ \cite{Altmannshofer}.
We have checked  that the SUSY contribution on the anomalous magnetic moment
is within the experimental allowed value  in all cases of Figures 2--6.

\begin{frame}{}
\begin{figure}[tb]
\begin{minipage}[]{0.45\linewidth} 
\hspace{-10mm}
{\includegraphics[width=7cm]{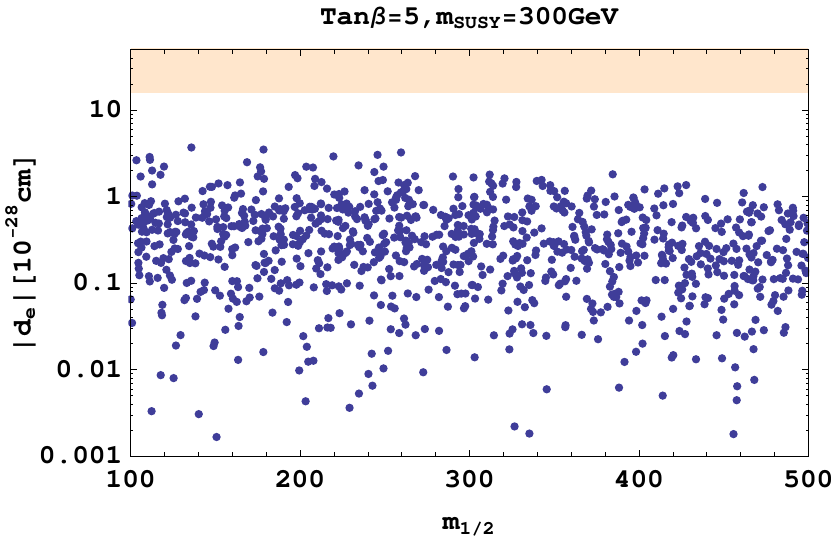}}
\caption{Electric dipole moment of the electron versus
 gaugino mass. 
The current experimental bound is $1.6\times 10^{-27}$[$e$cm]. }
\end{minipage}
\qquad
\begin{minipage}[]{0.45\linewidth} 
\hspace{-10mm}
{\includegraphics[width=7cm]{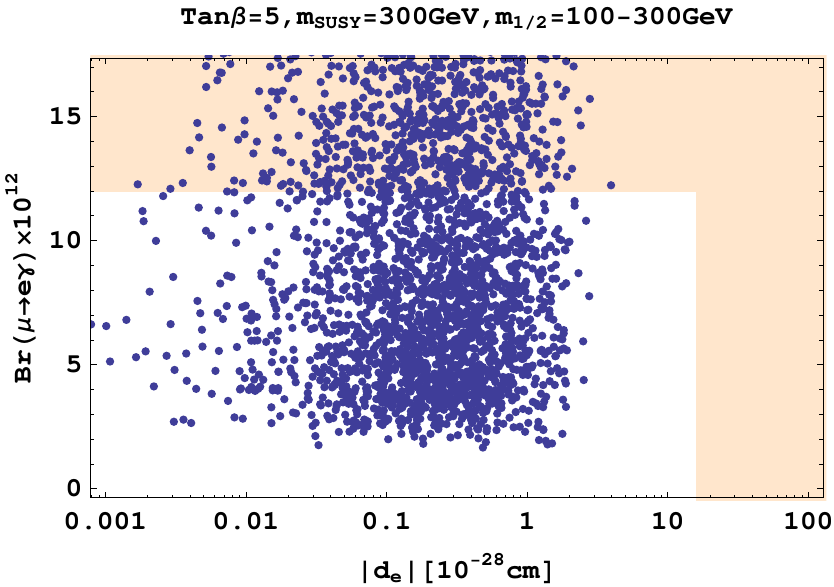}}
\caption{The branching ratio 
$\mu\rightarrow e\gamma$ versus the electric dipole moment the electron,  
  where $m_{\text{SUSY}}=300$GeV, $m_{1/2}=100-300$GeV and $\tan\beta=5$.}
\end{minipage}
\end{figure}
\end{frame}

\begin{figure}[t]
\subfigure[]{\includegraphics[width=7cm]{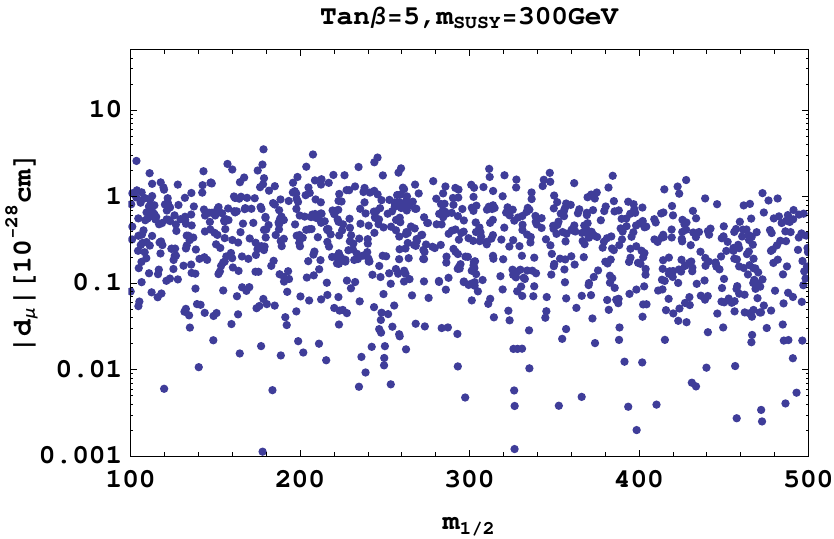}}
\qquad
\subfigure[]{\includegraphics[width=7cm]{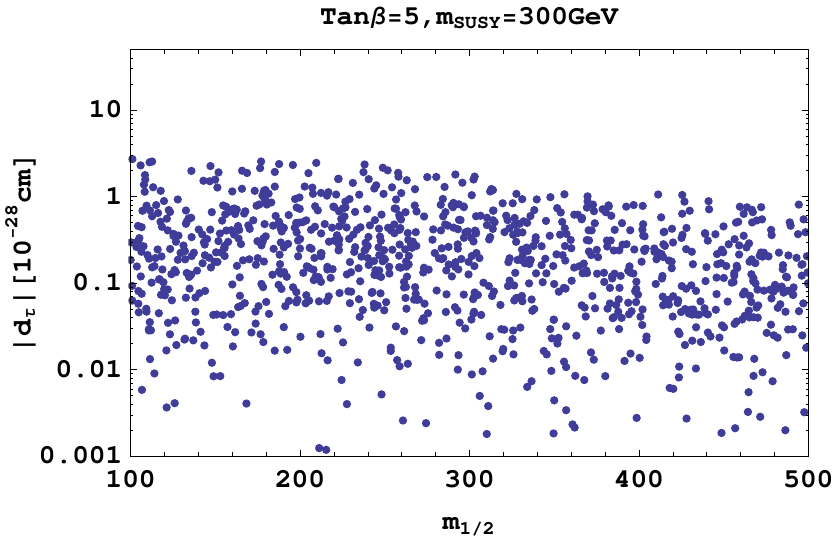}}
\caption{Electric dipole moments of (a) muon  and (b) tau,
 where  $m_{\text{SUSY}}=300$GeV, $m_{1/2}=100-500$GeV and
$\tan\beta=5$.}
\end{figure}

\section{Summary}

There appear  many flavor models with the non-Abelian discrete symmetry
 within the framework of SUSY. The flavor symmetry controls
the squark and slepton mass matrices 
as well as the quark and lepton mass matrices.
Therefore, the flavor models could be tested in the 
squark and slepton sectors.
We have discussed  slepton mass matrices 
in the  $S_4$ flavor model with SUSY $SU(5)$ GUT.
By considering the gravity mediation within the 
framework of supergravity theory, 
we have estimated the SUSY breaking in the slepton mass matrices,
which give the prediction for  the  $\mu \rightarrow e + \gamma$ decay
 and the electron EDM.

By taking  Yukawa couplings  to be in the region of 
$0.1$ to $1$ without tuning, we have obtained 
a lower bound for the ratio of $\mu\rightarrow e\gamma$ as $10^{-13}$
 if  $m_{\text{SUSY}}$ and  $m_{1/2}$ are below $500$GeV.
This predicted value  will be testable at the MEG experiment.
 The off diagonal terms of slepton mass matrices,
which come from the SUSY breaking, also contribute to   EDM of leptons.
 The natural prediction  of the electron EDM is around $10^{-29}-10^{-28}e$cm,
which can be tested by future experiments.
In our calculation,  we take $\Lambda$  to be the GUT scale.
Our predicted values crucially depend on 
$a_i=\langle \chi_i \rangle/\Lambda$, but not $\Lambda$.
 Since magnitudes of $a_i$ are determined by 
quark and lepton masses, our predictions are not changed even if
the $S_4$ scale $\Lambda$ is taken to be much larger or smaller than
the GUT one.

As shown in this work, the SUSY sector provides us  rich fields of
investigating flavor models with the non-Abelian discrete symmetry.

\vspace{1cm}
\noindent
{\bf Acknowledgement}

We owe the $S_4$ flavor model to Y. Shimizu.
H.I. is supported by Grand-in-Aid for Scientific Research,
No.21.5817 from the Japan Society of Promotion of Science.
The work of M.T. is  supported by the
Grant-in-Aid for Science Research, No. 21340055,
from the Ministry of Education, Culture,
Sports, Science and Technology of Japan.

\appendix

\section{Multiplication rule of $S_4$}

The $S_4$ group has 24 distinct elements and irreducible representations 
${\bf 1},~{\bf 1}',~{\bf 2},~{\bf 3}$, and ${\bf 3}'$.
The multiplication rule depends on the basis.
One can see its  basis dependence in our  review~\cite{Ishimori:2010au}.
We present  the multiplication rule, 
which is used in this paper:
\begin{align}
\begin{pmatrix}
a_1 \\
a_2
\end{pmatrix}_{\bf 2} \otimes  \begin{pmatrix}
                                      b_1 \\
                                      b_2
                                  \end{pmatrix}_{\bf 2}
 &= (a_1b_1+a_2b_2)_{{\bf 1}}  \oplus (-a_1b_2+a_2b_1)_{{\bf 1}'} 
  \oplus  \begin{pmatrix}
             a_1b_2+a_2b_1 \\
             a_1b_1-a_2b_2
         \end{pmatrix}_{{\bf 2}\ ,} \\
\begin{pmatrix}
a_1 \\
a_2
\end{pmatrix}_{\bf 2} \otimes  \begin{pmatrix}
                                      b_1 \\
                                      b_2 \\
                                      b_3
                                  \end{pmatrix}_{{\bf 3}}
 &= \begin{pmatrix}
          a_2b_1 \\
          -\frac{1}{2}(\sqrt 3a_1b_2+a_2b_2) \\
          \frac{1}{2}(\sqrt 3a_1b_3-a_2b_3)
      \end{pmatrix}_{{\bf 3}} \oplus \begin{pmatrix}
                                        a_1b_1 \\
                                        \frac{1}{2}(\sqrt 3a_2b_2-a_1b_2) 
\\
                                        -\frac{1}{2}(\sqrt 3a_2b_3+a_1b_3)
                                   \end{pmatrix}_{{\bf 3}'\ ,} \\
\begin{pmatrix}
a_1 \\
a_2
\end{pmatrix}_{\bf 2} \otimes  \begin{pmatrix}
                                      b_1 \\
                                      b_2 \\
                                      b_3
                                  \end{pmatrix}_{{\bf 3}'}
&= \begin{pmatrix}
         a_1b_1 \\
         \frac{1}{2}(\sqrt 3a_2b_2-a_1b_2) \\
         -\frac{1}{2}(\sqrt 3a_2b_3+a_1b_3)
     \end{pmatrix}_{{\bf 3}} \oplus
      \begin{pmatrix}
                                      a_2b_1 \\
                                      -\frac{1}{2}(\sqrt 3a_1b_2+a_2b_2) \\
                                      \frac{1}{2}(\sqrt 3a_1b_3-a_2b_3)
                                  \end{pmatrix}_{{\bf 3}'\ ,} \\
\begin{pmatrix}
a_1 \\
a_2 \\
a_3
\end{pmatrix}_{{\bf 3}} \otimes  \begin{pmatrix}
                                      b_1 \\
                                      b_2 \\
                                      b_3
                                  \end{pmatrix}_{{\bf 3}}
 &= (a_1b_1+a_2b_2+a_3b_3)_{{\bf 1}} 
  \oplus \begin{pmatrix}
             \frac{1}{\sqrt 2}(a_2b_2-a_3b_3) \\                                            
             \frac{1}{\sqrt 6}(-2a_1b_1+a_2b_2+a_3b_3)
         \end{pmatrix}_{\bf 2} \nonumber \\
 &\ \oplus \begin{pmatrix}
            a_2b_3+a_3b_2 \\
            a_1b_3+a_3b_1 \\
            a_1b_2+a_2b_1
         \end{pmatrix}_{{\bf 3}} \oplus \begin{pmatrix}
                                          a_3b_2-a_2b_3 \\
                                          a_1b_3-a_3b_1 \\
                                          a_2b_1-a_1b_2
                                       \end{pmatrix}_{{\bf 3}'\ ,} \\
\begin{pmatrix}
a_1 \\
a_2 \\
a_3
\end{pmatrix}_{{\bf 3}'} \otimes  \begin{pmatrix}
                                      b_1 \\
                                      b_2 \\
                                      b_3
                                  \end{pmatrix}_{{\bf 3}'}
 &= (a_1b_1+a_2b_2+a_3b_3)_{{\bf 1}}
  \oplus \begin{pmatrix}
             \frac{1}{\sqrt 2}(a_2b_2-a_3b_3) \\                                            
             \frac{1}{\sqrt 6}(-2a_1b_1+a_2b_2+a_3b_3)
         \end{pmatrix}_{\bf 2} \nonumber \\
 &\ \oplus \begin{pmatrix}
            a_2b_3+a_3b_2 \\
            a_1b_3+a_3b_1 \\
            a_1b_2+a_2b_1
         \end{pmatrix}_{{\bf 3}} \oplus \begin{pmatrix}
                                          a_3b_2-a_2b_3 \\
                                          a_1b_3-a_3b_1 \\
                                          a_2b_1-a_1b_2
                                       \end{pmatrix}_{{\bf 3}'\ ,} \\
\begin{pmatrix}
a_1 \\
a_2 \\
a_3
\end{pmatrix}_{{\bf 3}} \otimes  \begin{pmatrix}
                                      b_1 \\
                                      b_2 \\
                                      b_3
                                  \end{pmatrix}_{{\bf 3}'}
 &= (a_1b_1+a_2b_2+a_3b_3)_{{\bf 1}'}  
 \oplus \begin{pmatrix}
             \frac{1}{\sqrt 6}(2a_1b_1-a_2b_2-a_3b_3) \\
             \frac{1}{\sqrt 2}(a_2b_2-a_3b_3)
         \end{pmatrix}_{\bf 2} \nonumber \\
 &\ \oplus \begin{pmatrix}
            a_3b_2-a_2b_3 \\
            a_1b_3-a_3b_1 \\
            a_2b_1-a_1b_2
         \end{pmatrix}_{{\bf 3}} \oplus \begin{pmatrix}
                                          a_2b_3+a_3b_2 \\
                                          a_1b_3+a_3b_1 \\
                                          a_1b_2+a_2b_1
                                       \end{pmatrix}_{{\bf 3}'\ .}
\end{align}

 More details are shown in the review~\cite{Ishimori:2010au}.


\end{document}